\documentclass{emulateapj}
\usepackage{natbib}
\usepackage{graphicx}
\usepackage{appendix}
\usepackage{epsfig}
\usepackage{epsf}
\usepackage{amsmath}
\usepackage{amssymb}
\usepackage{url}
\usepackage{aas_macros}
\newcommand{\Hal}{\rm H\alpha}

\slugcomment{To appear in ApJ. XX.}
\shorttitle{PTF11kx at 3.5 Years}
\shortauthors{Graham et al.}

\begin{document}
\title{PTF11kx: A Type Ia Supernova with Hydrogen Emission Persisting After 3.5 Years}
\author{
M.~L. Graham$^{1,2}$, 
C.~E. Harris$^{2,3}$,
O.~D. Fox$^{4}$,
P.~E. Nugent$^{2,3}$, 
D.~Kasen$^{2,5}$,
J.~M. Silverman$^{6}$,
and A.~V. Filippenko$^{2}$
}

\altaffiltext{1}{Department of Astronomy, University of Washington, Box 351580, U.W., Seattle, WA 98195-1580, USA}
\altaffiltext{2}{Department of Astronomy, University of California, Berkeley, CA 94720-3411, USA}
\altaffiltext{3}{Lawrence Berkeley National Laboratory, 1 Cyclotron Road, MS 90R4000, Berkeley, CA 94720, USA}
\altaffiltext{4}{Space Telescope Science Institute, 3700 San Martin Drive, Baltimore, MD 21218, USA}
\altaffiltext{5}{Department of Physics, University of California, Berkeley, CA 94720, USA}
\altaffiltext{6}{Department of Astronomy, University of Texas, Austin, TX 78712, USA}

\begin{abstract}

The optical transient PTF11kx exhibited both the characteristic spectral features of Type Ia supernovae (SNe\,Ia) and the signature of ejecta interacting with circumstellar material (CSM) containing hydrogen, indicating the presence of a nondegenerate companion. We present an optical spectrum at $1342$ days after peak from Keck Observatory, in which the broad component of H$\alpha$ emission persists with a similar profile as in early-time observations. We also present {\it Spitzer} IRAC detections obtained $1237$ and $1818$ days after peak, and an upper limit from {\it HST} ultraviolet imaging at $2133$ days. We interpret our late-time observations in context with published results -- and reinterpret the early-time observations -- in order to constrain the CSM's physical parameters and compare to theoretical predictions for recurrent nova systems. We find that the CSM's radial extent may be several times the distance between the star and the CSM's inner edge, and that the CSM column density may be two orders of magnitude lower than previous estimates. We show that the H$\alpha$ luminosity decline is similar to other SNe with CSM interaction, and demonstrate how our infrared photometry is evidence for newly formed, collisionally heated dust. We create a model for PTF11kx's late-time CSM interaction and find that X-ray reprocessing by photoionization and recombination cannot reproduce the observed H$\alpha$ luminosity, suggesting that the X-rays are thermalized and that H$\alpha$ radiates from collisional excitation. Finally, we discuss the implications of our results regarding the progenitor scenario and the geometric properties of the CSM for the PTF11kx system.

\end{abstract}

\section{Introduction} \label{sec:intro}

Supernovae of Type Ia (SNe\,Ia) are thermonuclear explosions of carbon-oxygen white dwarf stars, and valuable for dark-energy cosmology studies as standardizable candles, but their progenitor scenario and their explosion mechanism are not yet well understood (e.g., \citealt{2011NatCo...2E.350H}). The two leading progenitor hypotheses are the double-degenerate (DD) scenario of two white dwarf stars, and the single-degenerate (SD) scenario of one white dwarf with either a red giant or main-sequence companion. In 2011, two nearby SNe\,Ia, SN\,2011fe and PTF11kx, were observed in such detail that strong constraints could be placed on their progenitor systems. For SN\,2011fe, the early-time light curve exhibited no interaction between SN ejecta and a companion star \citep{2011Natur.480..344N,2012ApJ...744L..17B}. Archival {\it Hubble Space Telescope (HST)} images excluded bright red giants and most main-sequence helium stars as potential companions \citep{2011Natur.480..348L} and ruled out a pre-existing supersoft X-ray source at its location \citep{2014MNRAS.442L..28G}. Observations support the double-degenerate scenario for SN\,2011fe in particular, and larger-sample studies eliminate single-degenerate companions for most SNe\,Ia (e.g., \citealt{2011ApJ...741...20B}).

In contrast, PTF11kx exhibited clear signs of the SN\,Ia ejecta interacting with multiple components of circumstellar material (CSM): at $\sim30$ days after peak brightness, deep \ion{Ca}{2} absorption lines transitioned to emission and a broad H$\alpha$ emission feature emerged, a clear sign of SN ejecta interacting with a hydrogen-rich CSM at a radius of $\sim 10^{16}$ $\rm cm$ from the progenitor, and a unique and unambiguous indication of the single-degenerate scenario \citep[][hereafter D12]{2012Sci...337..942D}. At 280 days past peak, PTF11kx remained $\sim3$ mag brighter than a typical SN\,Ia owing to the additional luminosity from the CSM interaction (D12). \citet[hereafter S13]{2013ApJ...772..125S} extended the analysis of PTF11kx to 680 days after peak brightness, showing that the late-time spectra are dominated by the effects of this CSM interaction (i.e., SN\,Ia-CSM) as the total luminosity declines. These studies conclude that the characteristics of PTF11kx are most consistent with a symbiotic binary progenitor system like RS Oph that experienced multiple past nova eruptions during the mass-accretion phase.

Together, SN\,2011fe and PTF11kx established that multiple progenitor pathways exist for SNe\,Ia, and PTF11kx remains the best observed and most clear-cut case of a SN\,Ia with a hydrogen-rich CSM. In this paper, we extend the analysis of PTF11kx from 680 to an unprecedented 1342 days after peak brightness with an optical spectrum from Keck Observatory, in which we observe the persistence of the broad H$\alpha$ emission line. We also present multi-band infrared (IR) imaging from the {\it Spitzer Space Telescope (Spitzer)} at $1237$ and $1818$ days, in which PTF11kx is detected at $3.6$ and $4.4$ $\mu$m. These observations are presented and analyzed in \S~\ref{sec:obs}. In \S~\ref{sec:disc}, we provide further discussion and constraints on the physical parameters of the CSM in the PTF11kx system, before concluding in \S~\ref{sec:con}. All dates are given in UT.

\section{Observations and Analysis}\label{sec:obs}

PTF11kx was discovered by the Palomar Transient Factory \citep{2009PASP..121.1334R,2009PASP..121.1395L} at right ascension $\alpha= 08^{\rm h}09^{\rm m}12.866^{\rm s}$ and declination $\delta= +46^\circ 18' 48.81''$ (J2000), in an anonymous spiral host ($\alpha= 08^{\rm h}09^{\rm m}12.95^{\rm s}$, $\delta= +46^\circ 18' 46.7''$) that is spectroscopically identified in the Sloan Digital Sky Survey (SDSS) DR8 as a star-forming galaxy at redshift $z=0.0466$ \citep{2011ApJS..193...29A}. PTF11kx reached peak optical brightness on 2011-01-29 (D12), at which time the spectral features typical of SNe\,Ia were clearly seen. These features were consistent with the overluminous subtype group similar to SN\,1991T \citep[e.g.,][]{1992ApJ...384L..15F}. The subtype of 91T-like SNe\,Ia has been connected with SNe\,Ia-CSM through a robust statistical analysis by \cite{2015A&A...574A..61L}, who show that this association is not the result of a luminosity bias and propose that the true physical origin of 91T-like SNe\,Ia is binary systems with a nondegenerate companion.

As a significant amount of the SN optical emission could be seen through the CSM, D12 point out that this means the CSM interaction was either of low optical depth or did not fully cover the disk of the photosphere -- and that since the early-time \ion{Ca}{2} absorption features were saturated, the former must be the case. Assuming the CSM is distributed in spherical shells, D12 estimate the total CSM mass from their observed \ion{Ca}{2} column density, finding it to exceed 5 M$_{\odot}$, and remark that this is a very large fraction of the total system mass. Instead, they infer that the CSM must be distributed in clumps or rings, a geometry that is further supported by a large Balmer decrement (D12) and by the erratic narrow-line evolution in equivalent width (EW) presented by S13.

The analysis of PTF11kx is extended to late times by S13, who show how the broad H$\alpha$ emission-line flux continues to increase for $\sim300$ days, remains bright, and then begins to decline sometime between 500 and 600 days after peak brightness. They interpret the start of the decline of the broad component as the time at which the SN ejecta had overtaken the majority of the CSM. As SN\,Ia ejecta travel at $\sim10,000$ $\rm km\ s^{-1}$ and reach $\sim5\times10^{16}$ $\rm cm$ at $\sim500$ days, this indicates quite a thick CSM width of $\Delta R_{\rm CSM} \approx 4 R_{\rm CSM}$ (discussed also in \S~\ref{ssec:disc.ncsm}). In addition, they show that the broad H$\alpha$ line retains a consistent shape throughout these epochs (from day $\sim300$ onward, as shown in Fig. 2 of S13), and that it is asymmetric, appearing blueshifted by additional red-side flux absorption, a common effect of dust formation in SN ejecta that causes greater extinction for emission from the receding hemisphere.

In \S~\ref{ssec:obs.keck}, we describe our extremely late-time observations at Keck Observatory and compare with the late-time observations of S13. We present multi-band IR photometry from {\it Spitzer} in \S~\ref{ssec:obs.spitz} and ultraviolet (UV) imaging from {\it HST} in \S~\ref{ssec:obs.hst}.

\subsection{Late-Time Optical Spectroscopy}\label{ssec:obs.keck}

\begin{figure*}
\begin{center}
\includegraphics[width=7.0in]{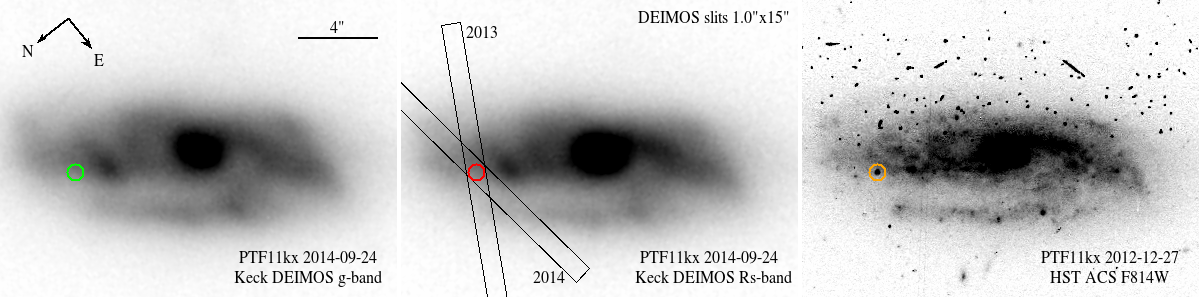}
\caption{Our Keck+LRIS images from 2014-09-24 (left and center) coregistered to a {\it HST}+ACS image from 2012-12-27 (right). The location of PTF11kx is marked with a circle in each image (green, Keck+LRIS $g$ filter; red, Keck+LRIS $R_s$ filter; and orange, {\it HST}+ACS F814W filter). At left, we have included a compass and a scale bar that applies to all images. At center, we have included the DEIMOS slit widths and orientations used for our 2014-10-02 spectrum presented in this paper, and the 2013-01-09 spectrum from S13 that we use here for comparison (in Figs. \ref{fig:2Dspec} and \ref{fig:comp}). At right, in the {\it HST}+ACS image, PTF11kx is located near a strip contaminated by cosmic rays that was not multiply-covered by the dither pattern of the CR-split exposures, but this does not affect our coregistration.  \label{fig:img}}
\end{center}
\end{figure*}

We obtained images of PTF11kx with the Low Resolution Imaging Spectrometer (LRIS; \citealt{1995PASP..107..375O}) at Keck Observatory on 2014-09-24: three 220~s exposures with the $g$ filter and three 180~s exposures with the $R_s$ filter. In Figure \ref{fig:img} we show these images, coregistered to an {\it HST} Advanced Camera for Surveys (ACS) image taken 2012-12-27 with the F814W filter (program GO-13024; PI J. Mulchaey). We do not see any source at the location of PTF11kx in either LRIS filter, and by comparing to SDSS catalogs we estimate the limiting magnitude at the location of PTF11kx in its host galaxy to be $g\gtrsim22$ mag.

\begin{figure*}
\begin{center}
\includegraphics[trim=0cm 0cm 0cm 0cm, clip=true, width=7.0in]{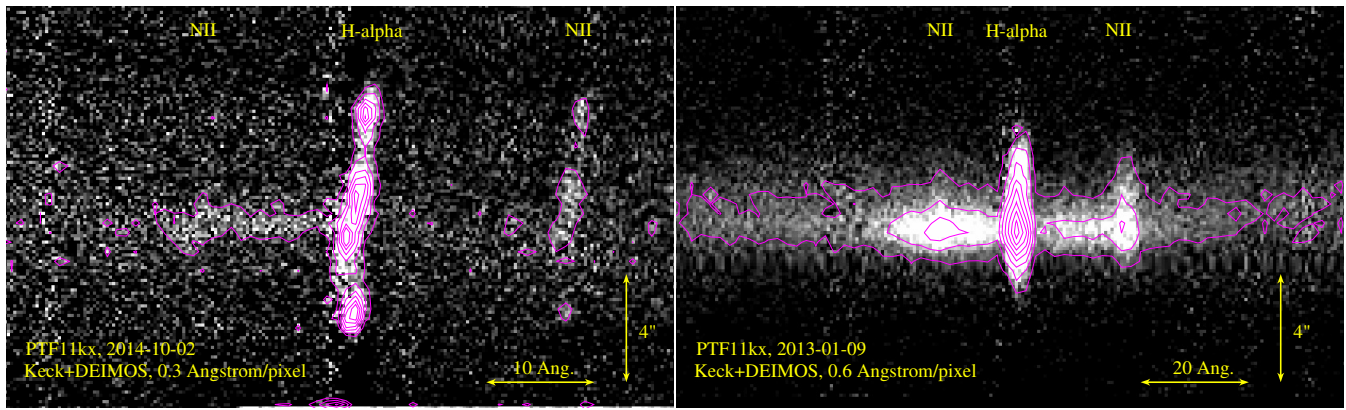}
\caption{Reduced and calibrated 2D DEIMOS spectra of PTF11kx in the region of H$\alpha$ and [\ion{N}{2}] from 2014-10-02 (left, using the 1200 lines $\rm mm^{-1}$ grating) and 2013-01-09 (right, using the 600 lines $\rm mm^{-1}$ grating; S13). We have used flux contours (pink) to highlight the broad component of H$\alpha$ (this especially helps for the 2014 spectrum at left) and the structure of the narrow H$\alpha$ feature. The narrow component of H$\alpha$ is clearly spatially extended at both epochs in a manner similar to [\ion{N}{2}], suggesting an association with the underlying host galaxy. In 2013 the pink contours show how the narrow lines peak at the same $y$-axis position as the trace of PTF11kx, suggesting a dominant contribution from the SN; however, by 2014 there is no discernible peak at the position of PTF11kx, suggesting the dominant contribution to the narrow features is from the host galaxy. This conclusion is independent of the difference in slit position angles shown in Figure \ref{fig:img}.  \label{fig:2Dspec}}
\end{center}
\end{figure*}

Despite this nondetection, we used an offset star to obtain a spectrum of PTF11kx with the DEep Imaging Multi-Object Spectrograph (DEIMOS; \citealt{2003SPIE.4841.1657F}) at Keck Observatory on 2014-10-02 (1342 days after peak brightness). We acquired three 900~s exposures using the 1200 lines $\rm mm^{-1}$ grating, the GG455 order-blocking filter, the 1.0\arcsec\ slit with a position angle of $96.8^\circ$ east of north, and a central wavelength of 6100 \AA. This configuration produces a scale of 0.33 \AA\ per pixel, a full width at half-maximum intensity (FWHM) resolution of 1.1--1.6 \AA, and wavelength coverage of 4500--7000 \AA. These spectra were reduced using routines written specifically for Keck LRIS/DEIMOS in the Carnegie {\sc Python} ({\sc CarPy}) package. The two-dimensional (2D) images were flat fielded, corrected for distortion along the $y$ axis (rectified), wavelength calibrated, cleaned of cosmic rays, and sky-subtracted. An example of one of our 2D spectra in the region of H$\alpha$ is shown in Figure \ref{fig:2Dspec}, in which we see narrow emission lines from a spatially extended source, and a broad component from a point source. We defined the apertures, extracted the one-dimensional (1D) spectra of the point source (i.e., PTF11kx), and combined them. Spectra of standard stars were used to determine the position, trace, and aperture size, as well as the sensitivity function. We deredden the spectrum for the effects of Milky Way extinction assuming $E(B-V)=0.052$~mag \citep{1998ApJ...500..525S}.

\begin{figure*}
\begin{center}
\includegraphics[trim=0.7cm 0cm 0.8cm 0cm, clip=true, width=3.4in]{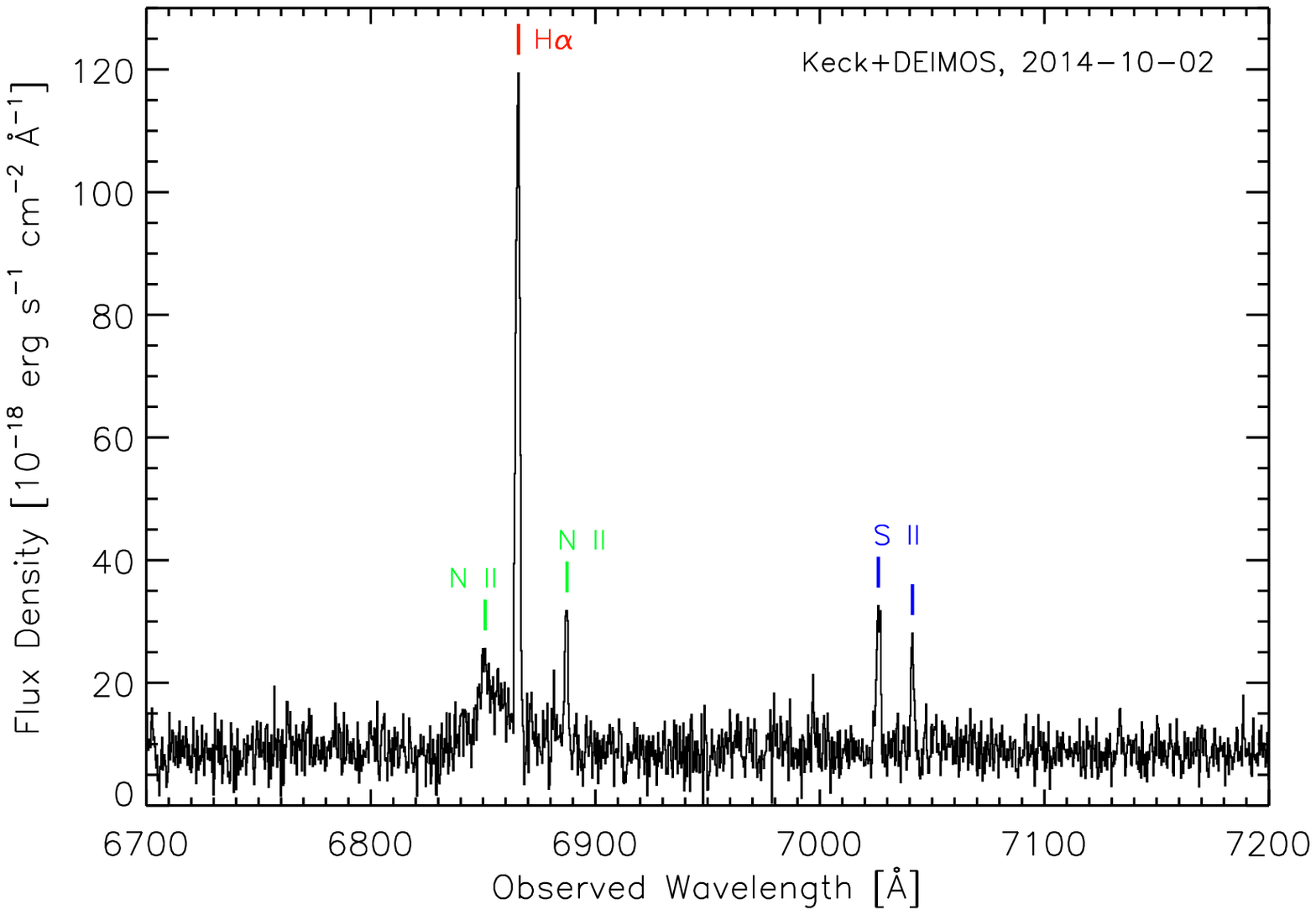}
\includegraphics[trim=0.8cm 0cm 0.8cm 0cm, clip=true, width=3.4in]{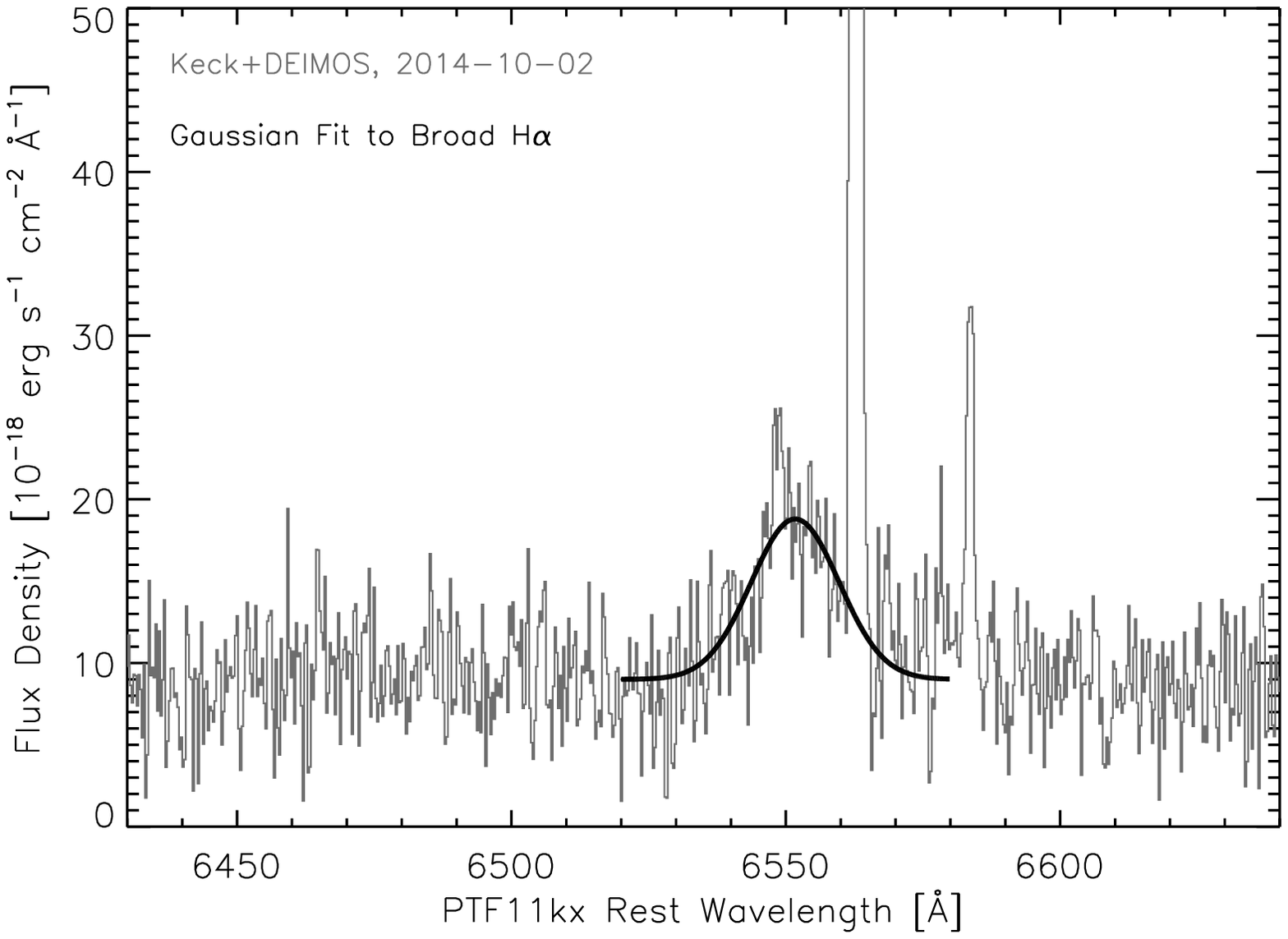}
\caption{Our Keck+DEIMOS spectrum from 2014-10-02. At left, we include a wide wavelength range and identify the narrow emission lines of H$\alpha$, [\ion{N}{2}], and [\ion{S}{2}] from the host galaxy that were used to obtain the relative velocity of PTF11kx. At right, we zoom in on the broad emission from H$\alpha$ and show the best-fit Gaussian function in the rest frame of PTF11kx. We describe the Gaussian fitting procedure in \S~\ref{ssec:obs.keck} and list the parameter values in Table \ref{tab:params}. \label{fig:1Dspec}}
\end{center}
\end{figure*}

The resulting 1D spectrum is shown in the left plot of Figure \ref{fig:1Dspec}, in observed wavelength for a limited wavelength range, outside of which the spectrum is mostly featureless. The trace of continuum emission from PTF11kx is not seen (e.g., Fig. \ref{fig:2Dspec}), and so we do not see the blue continuum flux increase blueward of $5700$ \AA\ as reported by S13. We do detect a broad, blueshifted H$\alpha$ emission line, as well as narrow features of H$\alpha$, H$\beta$, [\ion{N}{2}] $\lambda\lambda$6548, 6583, [\ion{S}{2}] $\lambda\lambda$6717, 6731, and [\ion{O}{3}] $\lambda$5007 (note that H$\beta$ and [\ion{O}{3}] are very faint, noisy features that we have opted not to plot, so they fall outside the wavelength range of Fig. \ref{fig:1Dspec}). We do not detect \ion {He}{1} $\lambda$5876 or $\lambda$7065, but S13 saw the former in their 680 day spectrum from 2013-01-09.

Based on the 2D version of our spectrum shown in Figure \ref{fig:2Dspec} and discussed below, the broad component of H$\alpha$ is from a point-like source because it is not spatially extended along the slit like the narrow features; the broad H$\alpha$ can be attributed entirely to the SN, but the narrow features are from underlying host-galaxy emission. Although the redshift of the host is $z=0.0466$, from the superposed narrow emission lines of H$\alpha$, [\ion{N}{2}], and [\ion{S}{2}] we determine that the correct redshift to apply is $z=0.0461\pm0.0001$. This is a velocity of $-150\ {\rm km\ s^{-1}}$ with respect to the host-galaxy's recession velocity and accounts for galaxy rotation, as PTF11kx is located on the approaching side of this inclined spiral (D12). We do not see any of the common nebular features from the underlying SN\,Ia (e.g., [\ion{Fe}{2}], [\ion{Fe}{3}], [\ion{Ni}{2}], or [\ion{Co}{3}]), but we would not expect to. In the $\sim1000$ day spectrum of normal SN\,Ia 2011fe the flux at $\lambda \approx 6500$ \AA\ was $\sim 5\times10^{-19}$ erg s$^{-1}$ cm$^{-1}$ \AA$^{-1}$ \citep{2015MNRAS.454.1948G}. At the distance of PTF11kx this would be $\sim 5\times10^{-22}$ erg s$^{-1}$ cm$^{-1}$ \AA$^{-1}$, far below the flux in the broad H$\alpha$ emission and essentially undetectable. In \S~\ref{sssec:obs.nar} and \S~\ref{sssec:obs.brd}, we interpret the narrow and broad components of H$\alpha$, respectively.

\subsubsection{The Narrow H$\alpha$ Feature}\label{sssec:obs.nar}

It is difficult to quantify the change in the narrow H$\alpha$ feature because it is contaminated by flux from the host galaxy. For example, the spatial distribution of the narrow emission along the slit (i.e., along the $y$ axis) in Figure \ref{fig:2Dspec}, as emphasized with the pink contour lines, shows no peak at the position of PTF11kx in 2014 like it did in 2013. This suggests that the narrow line is dominated by host emission in 2014, unlike earlier epochs when it is, in part, produced by the SN+CSM interaction (S13; D12). To further investigate the nature of the narrow-line evolution, for both the 2014 and 2013 epochs we measure three properties: (1) the ratio of the integrated, pseudocontinuum-subtracted fluxes in H$\alpha$ and [\ion{N}{2}], $F$(H$\alpha$)/$F$([\ion{N}{2}]); (2) the equivalent width of H$\alpha$, EW$({\rm H\alpha})$; and (3) the peak wavelength of H$\alpha$, $\lambda(\rm{H\alpha})$. For properties 1 and 2, the pseudocontinuum is determined from the regions immediately adjacent to, and on both sides of, each emission line. The results are  presented in Table \ref{tab:params2}.

\begin{table}
\begin{center}
\caption{Parameters of the narrow H$\alpha$ feature. \label{tab:params2}}
\begin{tabular}{lll} 
\hline
\hline
Date & Parameter & Value \\
\hline
2013-09-01  & $F$(H$\alpha$)/$F$([\ion{N}{2}])  & $4.5\pm0.9$ \\
                     & EW(H$\alpha$)                   & $15.9\pm3.2$ \AA \\
                     & $\lambda$(H$\alpha$)        & $6563.8\pm1.2$ \AA \\
2014-10-02  & $F$(H$\alpha$)/$F$([\ion{N}{2}])  & $6.0\pm1.2$ \\
                     & EW(H$\alpha$)                   & $16.2\pm3.2$ \AA \\
                     & $\lambda$(H$\alpha$)        & $6563.2\pm0.6$ \AA \\
\hline
\end{tabular}
\end{center}
\end{table}

In our narrow-line analysis, we find that the flux ratio $F$(H$\alpha$)/$F$([\ion{N}{2}]) increases from 2013 to 2014 (although still within 1$\sigma$ uncertainties), perhaps due to the change in relative contributions from the host and SN+CSM interaction to these lines. We find that the EW remains the same at $\sim16$~\AA, which is higher than the previously recorded low of $\sim7$~\AA\ at 393 days by S13. Since the flux from the underlying host galaxy is not expected to vary, this may suggest that there is still some narrow H$\alpha$ contribution from the SN in 2014 --- but it could also be attributed to a much higher SN continuum at 393 days. 

\begin{figure*}
\begin{center}
\includegraphics[trim=0.9cm 0.3cm 0.8cm 0cm, clip=true, width=5.4in]{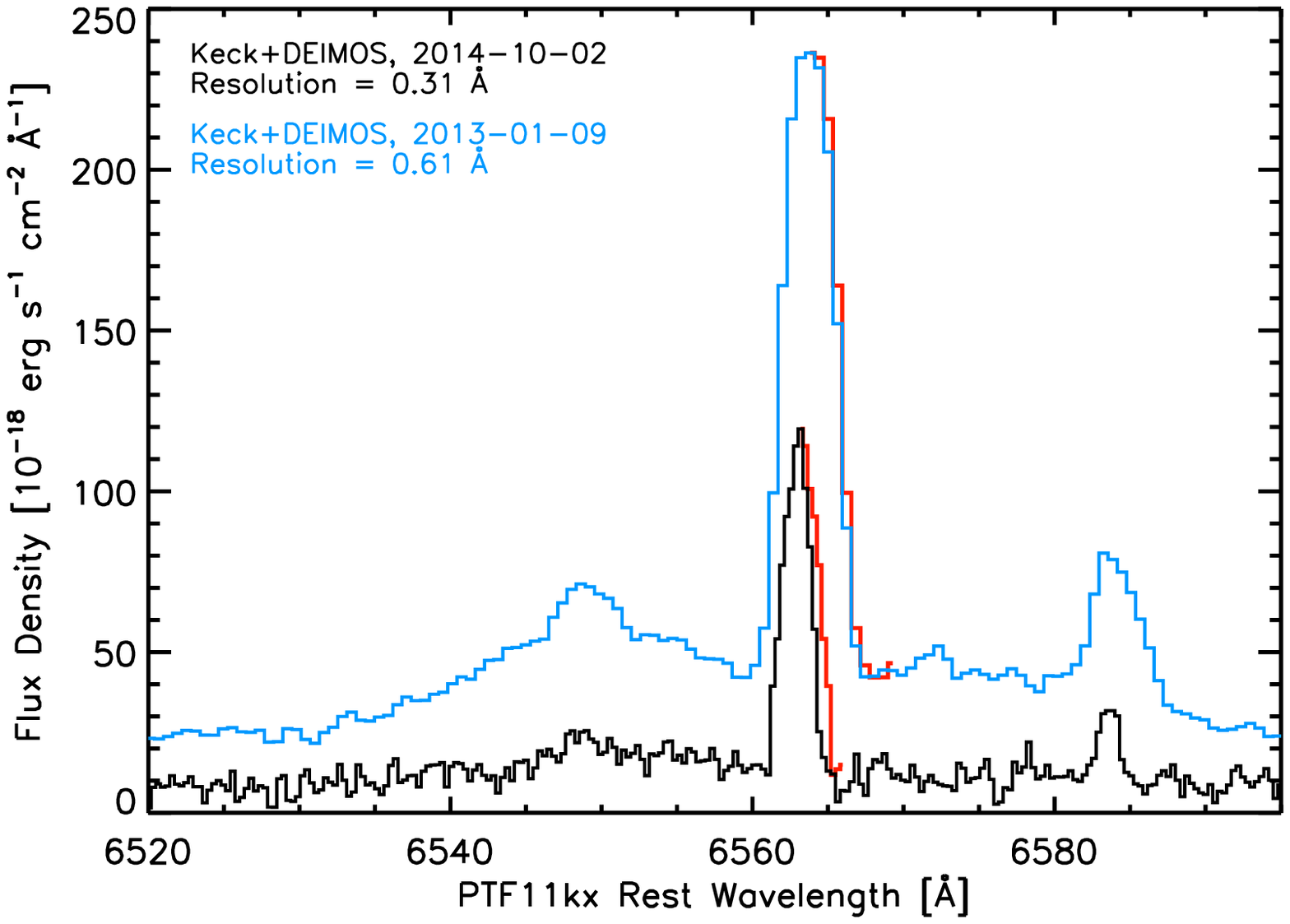}
\includegraphics[trim=3cm 1.5cm 6cm 2.5cm, clip=true, width=5.4in]{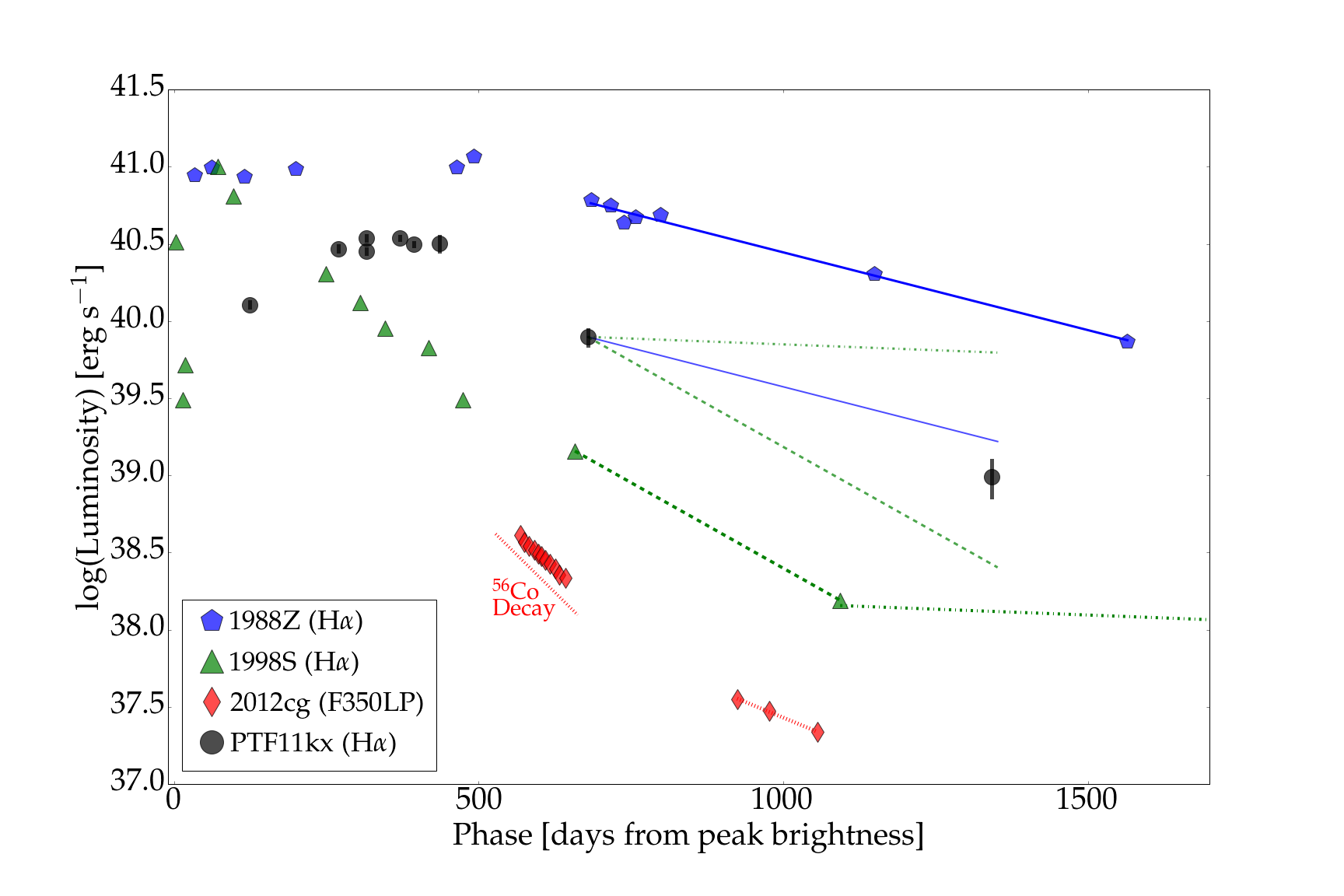}
\caption{A comparison of the H$\alpha$ in the PTF11kx spectrum on 2014-10-02 to past epochs. In the top panel we show the Keck+DEIMOS spectra from 2014-10-02 ($1342$ days, black, this work) and 2013-01-09 ($680$ days, blue; S13). The basic features are qualitatively the same at both epochs, but with lower flux at all wavelengths, and a less symmetric narrow H$\alpha$ line in 2014 (red lines are a reflection of the blue side of the narrow H$\alpha$). Narrow-line parameters are listed in Table \ref{tab:params2}. In the bottom panel we show the late-time evolution of the integrated flux of the broad component of H$\alpha$ for PTF11kx (black circles; phases $<1000$ days are from Fig. 2 of S13). For comparison, we also plot the H$\alpha$ luminosity for two SNe\,IIn with CSM interaction, SN\,1988Z (blue pentagons) and SN\,1998S (green triangles), with linear fits to their decline in log-luminosity (blue solid and green dashed lines). We have extended these lines from the $680$-day epoch of PTF11kx to demonstrate how its rate of decline fits into the broad range of decline rates exhibited by the CSM interaction of SNe. To compare with the expected decline luminosity of a normal SN\,Ia we have also included the late-time light curve of SN\,Ia 2012cg (red diamonds), with the decay rate of $^{56}$Co shown adjacent to the $<600$-day data (red dotted line) and a linear fit to the shallower decline rate at $\sim1000$ days (red dash-dot line). We do not extend these lines from the $680$-day observation of PTF11kx because we are emphasizing the comparison in luminosity, not decline rate, with SN\,2012cg. \label{fig:comp}}
\end{center}
\end{figure*}

Between 2013 and 2014 we find that the peak of the narrow H$\alpha$ line appears to have shifted $\sim 0.6$~\AA\ blueward; although this is within 1$\sigma$ uncertainties, we consider the physical implications of such a blueshift. By reflecting the blue side of the narrow H$\alpha$ onto the red side (see the thin grey lines in the left panel of Fig. \ref{fig:comp}), we show that the narrow H$\alpha$ is less symmetric in 2014 than in 2013; specifically, the red-side flux is depressed relative to the blue side, and this effect is more significant in 2014 than 2013. Such behavior is commonly attributed to the formation of dust in the system, because material that produces the redshifted emission is moving away from the observer and is on the far side of the system, and therefore has a larger line-of-sight column density of absorbing material. We do not expect an asymmetric narrow H$\alpha$ line if the emission is from the host galaxy alone, as there is not typically a direct correlation between line-of-sight distance and velocity of the emitting material. This asymmetry might be further evidence of the narrow H$\alpha$ line being formed at least in part by ongoing SN+CSM interaction, and not entirely dominated by host emission. However, a close examination of the pink contour lines in the 2014 2D spectrum (left-hand image in Fig. \ref{fig:2Dspec}) reveals a more likely explanation: the spatial structure over the extraction aperture has contrived to produce this asymmetry in 2014 compared to 2013, as the velocity profile appears dominated by spatially extended characteristics.

\subsubsection{The Broad H$\alpha$ Feature}\label{sssec:obs.brd}

Having determined that the narrow component of H$\alpha$ can be entirely attributed to host-galaxy emission, we now focus exclusively on the broad component. In the right plot of Figure \ref{fig:1Dspec} we show our 1D Keck+DEIMOS spectrum zoomed in on the broad H$\alpha$ feature, in the rest frame of PTF11kx, with a best-fit Gaussian to the broad emission. There are four pertinent details about this Gaussian fit: (1) we first fit and subtract the continuum, which was measured from either side of the H and [\ion{N}{2}] features; (2) we simultaneously fit a Gaussian to the [\ion{N}{2}] $\lambda$6548 line, holding the central wavelength fixed at 6548~\AA\ and the FWHM equal to that of the more isolated [\ion{N}{2}] $\lambda$6583 line, 1.6~\AA; (3) we fit only the data redward of the narrow H$\alpha$ feature; and (4) we ``bootstrap" our parameter uncertainties by permuting the flux uncertainties and refitting 1000 times, taking the final parameter values and uncertainties as the mean and standard deviation over all fits. The resulting Gaussian fit parameters and their uncertainties are listed in Table \ref{tab:params}.

\begin{table}
\begin{center}
\caption{Parameters of the Broad H$\alpha$ Feature. \label{tab:params}}
\begin{tabular}{ll} 
\hline
\hline
Parameter & Fit Value \\
\hline
Continuum $f_\lambda$    & $(9.0\pm2.9) \times 10^{-18}$ $\rm erg\ s^{-1}\ cm^{-2}\ \AA^{-1}$ \\
Peak $f_\lambda$             & $(9.8\pm0.9) \times 10^{-18}$ $\rm erg\ s^{-1}\ cm^{-2}\ \AA^{-1}$ \\
Integrated flux     & $(196\pm52) \times 10^{-18}$ $\rm erg\ s^{-1}\ cm^{-2}$ \\
Integrated luminosity  & $(2.7\pm0.7) \times 10^{5}$ L$_{\odot}$ \\
Equivalent width  & $38\pm15$ \AA \\ 
FWHM                  & $18.8\pm3.3$ \AA \\
                              & $860\pm150$ $\rm km\ s^{-1}$ \\
Peak wavelength  & $6552\pm1$ \AA \\
Line velocity          & $516\pm46$ $\rm km\ s^{-1}$\\
\hline
\end{tabular}
\end{center}
\end{table}

In Figure \ref{fig:comp} we assess the change in PTF11kx on 2014-10-02 since the last published spectrum, acquired on 2013-01-09, at 680 days after maximum brightness by S13. In the top plot, we compare the spectra directly and find that the basic features are qualitatively the same at both epochs, but with lower flux at all wavelengths in 2014. In particular, the broad H$\alpha$ feature has a similar width and blueshift, which suggests that it is produced by the same physical process at both epochs. 

In the bottom plot of Figure \ref{fig:comp} we add our observation at 1342 days to the data from S13 to show the evolution in luminosity of the broad component of H$\alpha$. As previously discussed, the start of the decline is interpreted as the shock front having crossed the extent of the CSM, and now we investigate how the rate of the decline might help us to physically constrain the CSM properties. It is impossible to characterize the decline in H$\alpha$ luminosity with only two points after $500$ days, so we instead compare with two Type IIn SNe having late-time spectroscopy. Although SNe\,IIn are driven by the core collapse of a massive star, it is an appropriate comparison in this case because, like PTF11kx, their emission is dominated by interaction between the CSM and the SN ejecta. From the six SNe\,IIn analyzed at $>3$~yr by \cite{2017MNRAS.466.3021S}, we chose to compare with SN\,1988Z and SN\,1998S because they represent systems with a higher H$\alpha$ luminosity and a steeper decline rate (SN\,1988Z) and vice versa (SN\,1998S). The H$\alpha$ integrated fluxes for SN\,1988Z were obtained from Table 4 of \cite{1999MNRAS.309..343A}, and corrected to luminosities using a distance of $70.7$ Mpc \citep{1998ApJ...500...51W}. The H$\alpha$ fluxes for SN\,1998S were obtained from \cite{2001MNRAS.325..907F}, \cite{2004MNRAS.352..457P}, and \cite{2012MNRAS.424.2659M}, and converted to luminosity using a distance of $15.5$ Mpc and accounting for a Galactic extinction with $A_V=0.68$ mag as in \cite{2012MNRAS.424.2659M}. The H$\alpha$ luminosity evolution for these SNe\,IIn is shown in Figure \ref{fig:comp}. For SN\,1998S, the slope represented by the green dot-dashed line extended from $1093$ days is calculated based on H$\alpha$ observations of $L_{\rm H\alpha} = 8.9$ and $3.7 \times 10^{37}$ $\rm erg\ s^{-1}$ at 2148 and 5097 days, respectively. We find that the decline rate of PTF11kx's H$\alpha$ luminosity is in between them, demonstrating that PTF11kx is consistent with other CSM-interaction dominated SNe at phases of $1000$ days.

In addition to being empirical specimens of two classes of observed SNe\,IIn, SN\,1988Z and SN\,1998S are representative of systems in which the emission is dominated by the forward or reverse shock, respectively. As described by \cite{1994ApJ...420..268C}, \cite{2012MNRAS.424.2659M}, \cite{2017MNRAS.466.3021S}, and others, SN ejecta first interact with the CSM released during the late stages of stellar evolution, when the mass-loss rate increased and shrouded the progenitor in a shell (or ring) of material, and so initially the emission is dominated by the forward shock of the SN through this denser material. After proceeding through the shell (or ring), the leading edge of the SN ejecta continues to interact with the mass lost during the earlier red supergiant (RSG) stage, when the mass-loss rate was lower and more continuous, and a reverse shock propagates backward through the denser CSM. At this later stage the emission becomes dominated by the reverse shock in the previously shocked dense ring rather than the forward shock in the newly shocked rarefied wind. The observational signature of this transition from forward- to reverse-shock dominated for SNe\,IIn is a rise in the relative emission strength of the oxygen, because the shell has become enriched with core-collapse SN ejecta material, and a shallower H$\alpha$ luminosity decline that is driven by ongoing interaction with RSG wind material. For SNe\,Ia the oxygen signature does not apply \citep{2015MNRAS.447..772F}, and we do not see any trace of an oxygen feature at $7300$~\AA. The question of whether the reverse or forward shock ``dominates" the H$\alpha$ emission is also slightly different for a SN\,Ia-CSM because there is no hydrogen in the unshocked ejecta component at radii interior to the reverse-shock front. For PTF11kx, ``reverse-shock dominated" would mean the H$\alpha$ luminosity is powered by X-ray and UV photons from the reverse shock, as we investigate in \S~\ref{ssec:revshock}, but the hydrogen itself is not necessarily near the reverse-shock boundary, as originally illustrated in Figure 2 of \citealt{1994ApJ...420..268C}. 

In Figure \ref{fig:comp} we also compare to the late-time luminosity of the normal SN\,Ia 2012cg, which did not exhibit any hydrogen features but may have had a nondegenerate companion star \citep{2016ApJ...820...92M}. These data are derived from {\it HST} WFC3 observations in the long-pass filter F350LP presented by \cite{2016ApJ...819...31G}, which demonstrated that the slope of the light curve at late times indicated a transition from being dominated by $^{56}$Co decay to having contributions from slower-decaying $^{57}$Co and other species. Although the $\sim1000$-day slope of SN\,2012cg is the closest match to the late-time decline rate of H$\alpha$ luminosity for PTF11kx, we find that the optical luminosity from radioactive decay is several orders of magnitude lower. However, the late-time optical luminosity of a SN\,Ia only represents decay energy trapped in the material and not, for example, that of gamma rays which are freely streaming after roughly day $100$. Could the H$\alpha$ in PTF11kx be excited by the gamma rays from radioactive decay? \cite{2013A&A...554A..67S} investigate the gamma-ray emission from the decay of $^{56}$Co and $^{57}$Co in SNe\,Ia, and we follow their formalism to estimate that the luminosity at $680$ and $1342$ days past peak would be $L_{\gamma} \approx 2 \times 10^{40}$ and $7 \times 10^{38}$ $\rm erg\ s^{-1}$, respectively (assuming a switch from $^{56}$Co to $^{57}$Co at 900 days). Although these values are consistent with the H$\alpha$ luminosity at $680$ and $1342$ days, most of the gamma rays will not interact with the CSM: the total attenuation cross sections for hydrogen in the presence of a SN\,Ia gamma-ray emission spectrum 0.1--3~MeV \citep{2013A&A...554A..67S} are $1.6\times10^{-30}$ to $5\times10^{-34}$ $\rm cm^{2}$ \citep{1973AD......5...51V}. When combined with an estimated maximum column density of $\sim10^{23}$ $\rm cm^{-2}$ (D12), we find that the optical depth of the CSM to gamma rays is $<<1$, and so we conclude that the broad H$\alpha$ luminosity is not powered by radioactive decay.

\subsection{Late-Time Infrared Imaging with {\it Spitzer}}\label{ssec:obs.spitz}

\begin{figure}
\begin{center}
\includegraphics[trim=0cm 1.5cm 0cm 1cm, clip=true, width=3.3in]{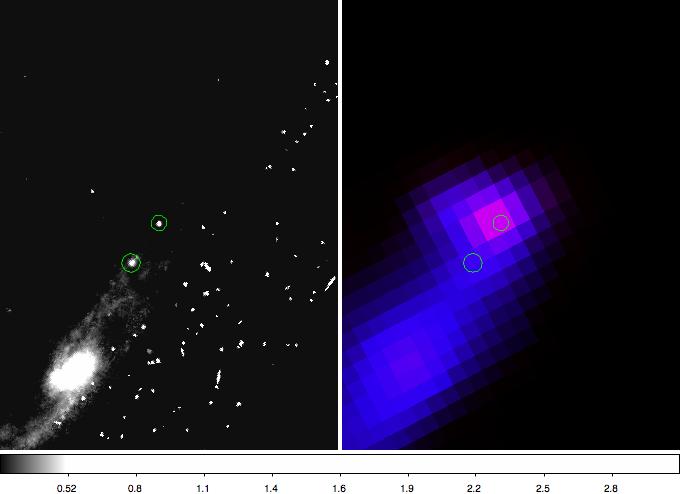}
\caption{{\it Spitzer} 3.6 $\mu$m image obtained on 2014-06-08 (right) coregistered with the {\it HST} image from Figure \ref{fig:img} (left). Circles mark PTF11kx (upper right) and a nearby bright region (lower left) to guide the eye. PTF11kx is clearly detected by {\it Spitzer} (pink) on top of the host-galaxy background (blue). \label{fig:IRimg}}
\end{center}
\end{figure}

We acquired two epochs of photometry with {\it Spitzer}'s Infrared Array Camera (IRAC; \citealt{2004ApJS..154...10F}) at 3.6 and 4.5 $\mu$m, summarized in Table \ref{tab:spitzer}.  The first epoch was obtained 105 days before our Keck+DEIMOS spectrum, and the second epoch 1.3~yr after. In Figure \ref{fig:IRimg} we show the {\it Spitzer} 3.6 $\mu$m image obtained at 1237 days, coregistered with an {\it HST} image of PTF11kx, and find IR emission at the location of PTF11kx that is securely detected above the host-galaxy background. These {\it Spitzer} observations suggest the presence of warm dust in the PTF11kx system; together with our optical observations we attempt to constrain the origin and heating mechanism of the dust, and distinguish whether it is newly formed or pre-existing.

\begin{table*}
\begin{center}
\caption{Spitzer Observations of PTF11kx. \label{tab:spitzer}}
\begin{tabular}{cccccccc} 
\hline
 & & & & & \multicolumn{3}{c}{ ------------ Dust Properties ------------ } \\
Phase & Date & Program ID & $3.6$ $\mu$m Flux & $4.5$ $\mu$m Flux & Mass & Temperature & Luminosity \\
(days) & (yyyy-mm-dd) & & (mJy) & (mJy) & (M$_{\odot}$) & (K) & (L$_{\odot}$) \\
\hline
1237 & 2014-06-08 & 11053 & $0.05\pm0.02$ & $0.06\pm0.03$ & $0.004$ & $587$ & $5.7 \times 10^7$ \\
1818 & 2016-01-10 & 10139 & $0.02\pm0.02$ & $0.03\pm0.03$ & $0.002$ & $584$ & $2.6 \times 10^7$ \\
\hline
\hline
\end{tabular}
\end{center}
\end{table*}

\begin{figure*}
\begin{center}
\includegraphics[trim=0.4cm 0.3cm 0.7cm 0.0cm, clip=true, width=3.5in]{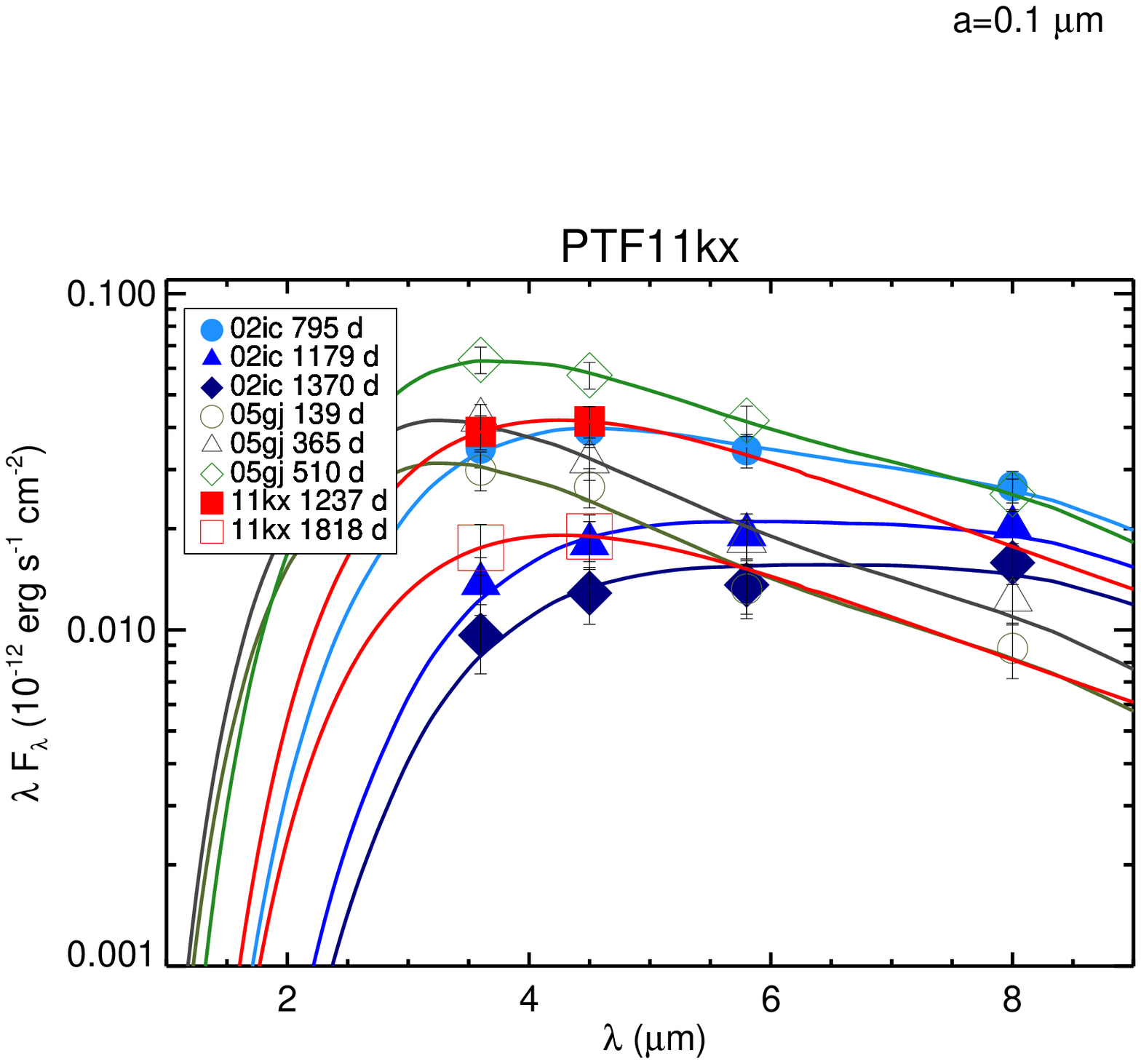}
\includegraphics[trim=0.4cm 0.0cm 0.8cm 0.5cm, clip=true, width=3.5in]{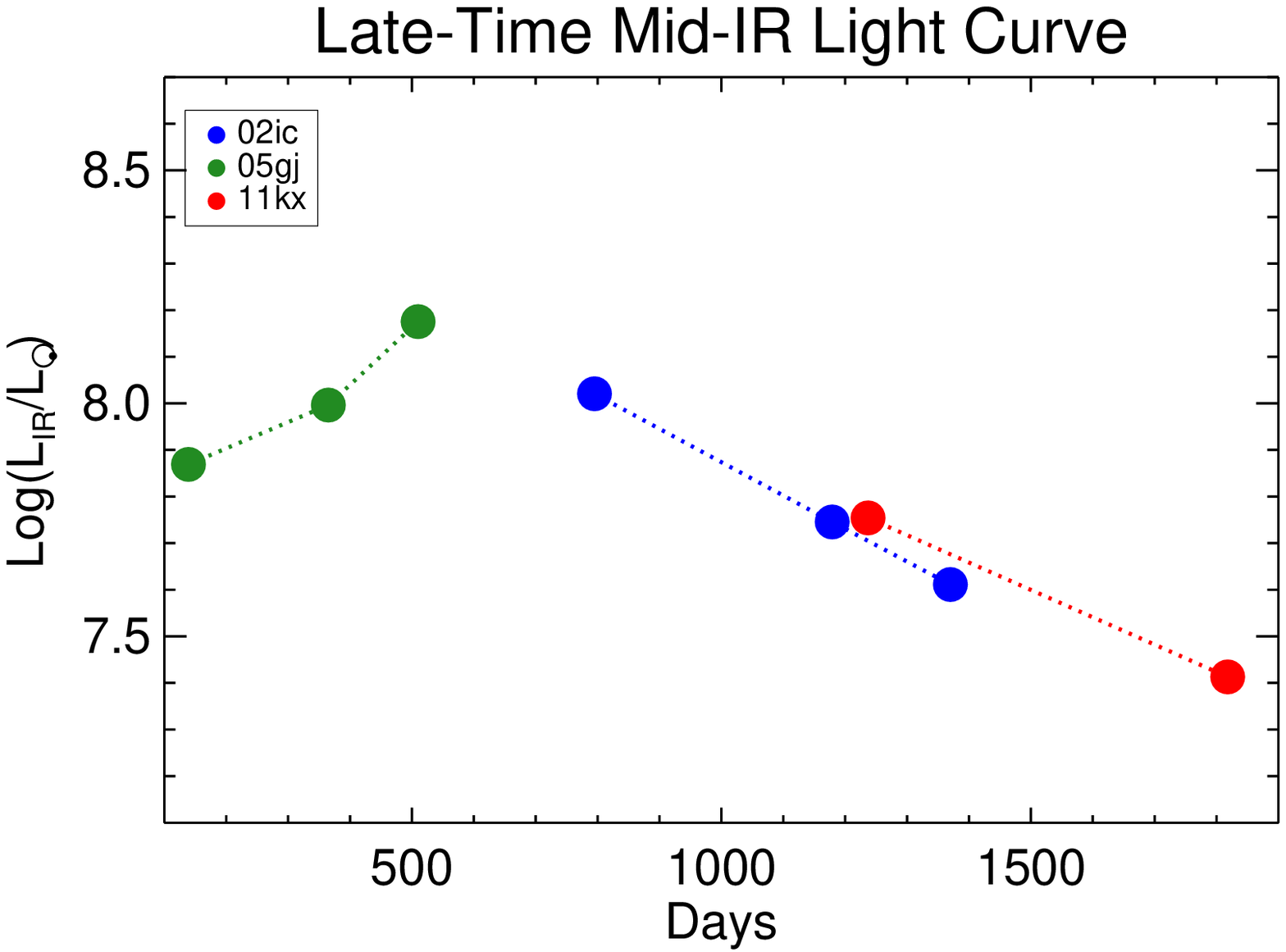}
\caption{ {\it Spitzer} observations at 3.6 and 4.5 $\mu$m of PTF11kx compared to SNe\,Ia-CSM 2002ic and 2005gj (and other SNe) from \protect\cite{2013ApJ...772L...6F}. At left, the IR SEDs for a variety of late-time phases are fit with blackbody curves. At right, the integrated IR luminosity evolution of PTF11kx at late times is similar to that of SNe\,Ia-CSM 2002ic and 2005gj. \label{fig:IRSED}}
\end{center}
\end{figure*}

In Figure \ref{fig:IRSED} we present the spectral energy distribution (SED) of PTF11kx, and compare it to other SNe\,Ia-CSM 2002ic and 2005gj from \cite{2013ApJ...772L...6F}. We find that the IR SED of PTF11kx has a shape and luminosity that is similar to those of these other two examples, and that its late-time decline in IR luminosity is similar to that of SN\,2002ic. We fit a blackbody to the IR SED and calculate a dust mass and temperature, which are also summarized in Table \ref{tab:spitzer}. As in \cite{2013ApJ...772L...6F}, we constrain the origin and heating mechanism of the dust by assuming a spherically symmetric, optically thin dust shell. This yields the blackbody radius for the dust of $R_{\rm dust}$, given by

\begin{equation}
R_{\rm dust} = \left( \frac{L_{\rm dust}}{4 \pi \sigma T_{\rm dust}^4} \right)^{1/2},
\end{equation}

\noindent
where $T_{\rm dust} = 587$ $\rm K$ is determined from a blackbody fit to the IR SED and $L_{\rm dust}$ is derived by integrating that blackbody, both of which are listed in Table \ref{tab:spitzer}, and $\sigma$ is the Stefan-Boltzmann constant. For the first epoch at $1237$ days we find that $R_{\rm dust} \approx 5 \times 10^{16}$~cm, or $\sim 0.015$~pc (note that this is a minimum estimate). For the second epoch  at $1818$ days we find that $R_{\rm dust}$~is the same within the error bars (i.e., the dust temperature is about the same and the flux uncertainties are $\sim100\%$). Since $R_{\rm dust} \approx 5 \times 10^{16}$ $\rm cm$ is consistent with the vaporization radius for a typical SN\,Ia peak luminosity of $10^{10}$ L$_{\odot}$ assuming 0.1~$\mu$m grains (as demonstrated by Fig. 8 of \citealt{2010ApJ...725.1768F}), it is more likely to be newly formed than pre-existing. However, we cannot rule out that (for example) significant clump-enabled self-shielding helped to preserve existing dust. Furthermore, we note that the blueshift in the broad H$\alpha$ line --- commonly interpreted as a signature of dust formation in the line-forming region --- was exhibited with the same peak wavelength since at least $124$ days after peak brightness (S13), which means that any dust created post-explosion started forming quite early.
 
To investigate the potential heating mechanism of the dust, we consider that this radius is also consistent with the extent of the CSM, $\sim5\times10^{16}$ $\rm cm$ (S13, and the second paragraph of \S~\ref{sec:obs}), and that even the bulk of the slower SN ejecta traveling at $\sim5000$ $\rm km\ s^{-1}$ has passed this distance by $1200$ days. This means that the emission from the CSM interaction is no longer dominated by the forward shock, as also discussed in \S~\ref{sssec:obs.brd}, which is the typical source of radiation to heat the dust in, for example, SNe\,IIn \citep{2013AJ....146....2F}. Furthermore, the inferred optical luminosity required to radiatively heat dust at the distance and temperature we observe for the PTF11kx system is $L_{\rm opt} \approx 10^8$ L$_{\odot}$. Assuming a bolometric absolute magnitude of $4.75$ for the Sun, we estimate that the optical apparent magnitude of the PTF11kx system would have to have been $\sim21.2$ mag. This is brighter than our estimated limiting magnitude of $g\gtrsim22$ at $1334$ days derived in \S~\ref{ssec:obs.keck}, suggesting that the heating mechanism is more likely to be collisional than radiative, which agrees with the fact that we have identified this dust as residing in the vicinity of the post-shock CSM. As an additional note, we point out that while some models of the post-shock companion star predict an increase in luminosity, it is typically $<10^4$ L$_{\odot}$, and in some cases the secondary actually becomes significantly fainter \citep{2003astro.ph..3660P,2012ApJ...760...21P,2013ApJ...773...49P,2016PASJ...68...11N}. Even the maximum possible increase in luminosity seen in these models is too low for the dust to be heated by emission from the shocked companion.

Therefore, unlike most SNe\,IIn and SNe\,Ia-CSM 2002ic and 2005gj, in which a massive pre-existing dust shell is radiatively heated by the forward shock \citep{2013AJ....146....2F,2013ApJ...772L...6F}, we conclude that the dust in PTF11kx is likely to be newly formed and collisionally heated, although we cannot rule out contributions from pre-existing dust and radiative heating (e.g., from the reverse shock). In terms of dust formation, PTF11kx may be similar to the Type Iax SN 2014dt, for which a late-time IR excess indicated $10^{-5}$ M$_{\odot}$ of newly formed dust, but pre-existing dust from the nondegenerate companion could not be ruled out \citep{2016ApJ...816L..13F}.

\subsection{Late-Time Ultraviolet Imaging with {\it HST}}\label{ssec:obs.hst}

We acquired a single epoch of UV imaging with {\it HST}'s Wide Field Camera 3 (WFC3) under Cycle 24 snapshot program GO-14779 (PI M. L. Graham). This program is designed to look for the UV signature of late-onset CSM interaction in a large sample of SNe\,Ia. Most of the objects targeted by this survey were chosen to be $1$--$3$~yr past peak brightness in Cycle 24. Despite its advanced age, we included PTF11kx because of its history of energetic CSM interaction. The images were obtained on 2016-12-01 UT ($2133$ days past peak brightness) with the F275W filter and have a total integration time of $858$~s, which we split into two exposures to mitigate contamination by cosmic rays. We use the pipeline processed ``{\sc drc}'' images, which have been drizzled, corrected for charge-transfer efficiency, and combined. 

\begin{figure}
\begin{center}
\includegraphics[trim=0cm 0cm 0cm 0cm, clip=true, width=3.3in]{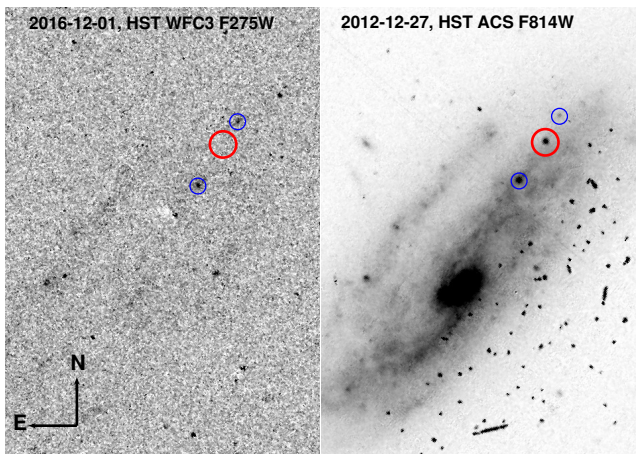}
\caption{At left we show our UV {\it HST} imaging from 2016 December 1, in which PTF11kx is undetected at $2133$ days past peak brightness. At right, the optical {\it HST} image in which PTF11kx is visible $698$ days after peak brightness (from Fig. \ref{fig:img}). Red 0.5\arcsec\ circles indicate the location of PTF11kx, and smaller blue circles mark nearby regions visible in both epochs and filters to guide the eye. \label{fig:hst}}
\end{center}
\end{figure}

In Figure \ref{fig:hst} we compare our day $2133$ UV image to the day $698$ optical {\it HST} image shown also in Figures \ref{fig:img} and \ref{fig:IRimg}. No point source is seen at the location of PTF11kx in its host galaxy. To estimate an upper limit for our observation we use a forced flux measurement with an aperture of $r=0.4\arcsec$, which encloses $86.2\%$ of the flux, and the AB zeropoint for this aperture of $24.0$~mag\footnote{As instructed in the WFC3 Data Handbook (\url{http://www.stsci.edu/hst/wfc3}), we use Tables 3 and 5 from \url{http://www.stsci.edu/hst/wfc3/documents/ISRs/WFC3-2009-31.pdf} for the $r=0.4\arcsec$ aperture flux correction and zeropoint.}. Based on this analysis, we find a limiting magnitude of $m_{\rm F275W} > 23.5$ for PTF11kx at $2133$ days past peak brightness. We convert this to a luminosity and find an upper limit of $L_{\rm UV} < 10^7$ L$_{\odot}$, indicating that no new, late-onset, energetic CSM-interaction episode is occurring.

\section{Discussion}\label{sec:disc}

In this discussion we combine our analysis of our late-time observations with previously published data for PTF11kx and provide further assessment of the physical parameters of the system. We assume that the CSM is of solar metallicity, and that the fastest SN ejecta move with $v_{\rm ej} \approx 20,000$ $\rm km \ s^{-1}$. The appearance of narrow emission lines $\sim60$ days after explosion marks the time at which the fastest SN ejecta reach the innermost edge of the CSM, securely constraining the distance between the white dwarf and CSM to $R_{\rm CSM} \approx 10^{16}$ $\rm cm$ (D12). In \S~\ref{ssec:disc.model}, we begin with a summary of recurrent-nova models and their predictions for the mass and distribution of CSM in a progenitor system like PTF11kx, to which we will compare our derived physical parameters. In \S~\ref{ssec:disc.N}, we reinterpret the pre-impact observations of PTF11kx, finding that the CSM column density could be significantly lower than estimated by D12, and also constrain the radial extent and thus the particle number density of the CSM. Section \ref{ssec:revshock} examines whether the late-time H$\alpha$ emission could be driven by ionizing photons generated by the reverse shock. Section \ref{ssec:disc.ring} briefly discusses the CSM geometry, and in \S~\ref{ssec:disc.nond} we consider an alternative interpretation of interaction with a nondegenerate companion star itself, instead of its CSM.

\subsection{CSM Models for Recurrent-Nova Systems}\label{ssec:disc.model}
 
The mass and distribution of CSM inferred from observations of PTF11kx suggests that the progenitor was a white dwarf with a red giant companion in a recurrent-nova system. The physical picture is that there are at least two kinematically distinct regions of CSM (which we call ``shells'' though they are of unknown angular extent), with a slower shell lying much farther from the SN than the nearer shell. The evidence for this includes the multiple velocity components in its absorption features and the erratic behavior of the narrow-line emission (S13) from photoionized material beyond the shock front, which suggests the shock interaction is variable. The impact of the inner shell is seen, but not the second.

There are several models for recurrent-nova systems. \citet[hereafter MB12]{2012ApJ...761..182M} model the creation of CSM shells in a symbiotic recurrent-nova system, where a white dwarf accretes material from the wind of a donor star. In contrast, recurrent novae from cataclysmic variable systems involve mass accretion from a companion star overflowing its Roche lobe (i.e., a close binary system). The timescales, accretion efficiency, and total mass lost can be significantly different between these two types of systems; in particular, accretion from a wind is much less efficient and symbiotic systems can generate a higher CSM mass. In the 1D models of MB12, after the ejecta from a nova eruption have swept up the wind material and the shock has cooled and decelerated, the resulting CSM shell has a thickness of $\Delta R_{\rm CSM} \approx 0.1 \ R_{\rm CSM}$, where $R_{\rm CSM}$ is the distance from the progenitor star to the CSM. However, in that model, shells of material from separate outbursts stack to create a more extended CSM. \cite{2016ApJ...823..100H} find that a collision with concentric, kinematically distinct nova shells creates a light curve that is significantly longer than the sum of the light curves of two individual shells. We will keep in mind these models of CSM interaction as we proceed through the next sections with estimates of the CSM physical parameters.

\subsection{Estimating the CSM Particle Density}\label{ssec:disc.N}\label{ssec:disc.ncsm}

D12 use high-resolution spectra of PTF11kx obtained at peak brightness, before the onset of interaction, to measure the equivalent width (EW$_{\rm Ca~II}$) of the blueshifted, saturated \ion{Ca}{2}~H\&K lines ($\lambda_{\rm Ca~II,H}=3968.47$~\AA, $\lambda_{\rm Ca~II,K}=3933.66$~\AA), from which they estimate the column density of calcium to be $N_{\rm Ca~II} \approx  5\times10^{18}$ $\rm cm^{-2}$. They show that, assuming the CSM is of solar abundance and in a spherical shell of $R \approx 10^{16}$ $\rm cm$, this implies a total CSM mass of $M_{\rm CSM} \approx 5.36$ M$_{\odot}$. As this is much larger than expected for a recurrent-nova progenitor system, D12 infer that the CSM must be distributed in a ring geometry instead of a spherically symmetric shell. Since this is a significant conclusion regarding the progenitor system, we reassess this measure of column density from the \ion{Ca}{2} absorption feature in the pre-maximum spectra after reconsidering the appropriate regime on the curve of growth and incorporating turbulence effects, with additional analysis of the \ion{Na}{1}~D lines and the optical depth of the CSM (all of which is described in detail in Appendix \ref{ap.a}). We find that the column density of \ion{Ca}{2} could be as low as $N_{\rm Ca} \approx 10^{16}$ $\rm cm^{-2}$, more than two orders of magnitude lower than estimated by D12. Under the assumption of solar abundance (i.e., that the number ratio of Ca to H is approximately $1$ to $5\times10^5$), this is a total particle column density of $N_{\rm CSM} \approx 5 \times 10^{21}$ $\rm cm^{-2}$. 
 
In order to translate a column density into a particle volume density we must first estimate the radial extent of the CSM. The integrated luminosity of broad H$\alpha$ decreases steeply around 500 days after explosion, indicating the end of interaction (the end of energy deposition) -- i.e., that the shock has swept over the CSM. Using this observation and the inferred impact time of $\sim 40$ days after explosion, one can estimate the CSM extent from Equation 6 of \cite{2016ApJ...823..100H} to be $\Delta R_{\rm CSM} \approx 4 R_{\rm imp}$, where $R_{\rm imp}$ is the inner radius of the CSM.  This is in agreement with the estimate of S13, as presented in \S~\ref{sec:obs}. We note, however, that it neglects possible re-energization from a collision with the second shell \citep{2016ApJ...823..100H}.

We combine our revised CSM column density and this CSM depth to estimate that the CSM particle volume density is $n_{\rm CSM} = N_{\rm CSM}/\Delta R_{\rm CSM} \approx 1.3 \times 10^{5}$ $\rm cm^{-3}$. We consider $N_{\rm CSM} \approx 10^{23}$ $\rm cm^{-2}$, as estimated by D12, to be an upper limit that gives $n_{\rm CSM} \lesssim 1.0\times 10^{7}$ $\rm cm^{-3}$.

If we assume a solar abundance (average mass per atomic particle of $\sim1.3$ times the proton mass) and a spherical shell geometry for the CSM in the PTF11kx system, our revised estimates for the density and radial extent of the material lead to a total mass estimate of $M_{\rm CSM} \approx 0.06$ M$_{\odot}$. This value is significantly lower than the estimate of D12, by about two orders of magnitude, but similar to derived estimates of CSM mass for other SNe\,Ia such as Kepler \citep{2015ApJ...808...49K} and SN\,2003du \citep{2004ApJ...607..391G}. It is much larger than the amount of CSM estimated by mass build-up models ($2\times10^{-6}$ M$_{\odot}$, MB12) and may indicate a phase of rapid mass loss by the companion in the years before the white dwarf explodes. Together with the dust-mass estimate in Table \ref{tab:spitzer}, this implies a gas-to-dust ratio of $\sim15$ in the post-shock CSM of PTF11kx. This is lower than the typical ratio of $100$--$200$ for the CSM of stars undergoing mass loss (which is similar to the ratio for the interstellar medium), but not implausible as other SNe have been observed to form unexpectedly large dust masses (e.g., \citealt{2015ApJ...800...50M,2015MNRAS.446.2089W}).

\subsection{Reverse-Shock Ionization of the CSM}\label{ssec:revshock}

As previously discussed, we interpret the late-time observations of PTF11kx at $>500$ days as indication that the forward shock has passed over the bulk of the CSM and is no longer powering the broad $\Hal$ emission. However, the reverse shock lives on and continues to produce ionizing photons that propagate into the cooled, fast CSM. Thus, some of the observed broad $\Hal$ luminosity may be the result of continued photoionization and photo-recombination powered by the reverse shock. To investigate this, we evolve a model like those described by \cite{2016ApJ...823..100H} with an impact time of 50 days after explosion, CSM density $10^{-18}~{\rm g~cm^{-3}}$ (i.e., intermediate between our estimate and that of D12), and CSM depth $\Delta R = 4R_{\rm imp}$. This CSM has a mass of $\sim 0.3$ M$_\odot$, nearly the same as the outer ejecta, so the reverse shock enters the inner ejecta at about the time the forward shock overtakes the edge of the CSM. The SN ejecta profile has the same properties as in \cite{2016ApJ...823..100H}, and for these CSM properties its outer edge velocity is at $\sim 20,000~{\rm km~s^{-1}}$.

To estimate the plausibility that the reverse shock powers $\Hal$ through photoionization equilibrium, here we naively estimate that one $\Hal$ photon is produced per ionizing photon absorbed. Then the broad $\Hal$ luminosity is related to the ionizing photon production rate, $\dot Q_{\rm ion}$, simply by

\begin{equation}\label{eq:LHaI}
	L_{\Hal} = E_{\Hal} \dot Q_{\rm ion},
\end{equation}

\noindent where $E_{\Hal}$ is the energy of an $\Hal$ photon. The ionizing photon production rate is defined by

\begin{equation}\label{eq:Qion}
\dot Q_{\rm ion}=\int ^{\nu_{\rm abs}} _{{\rm Ry}/h}\dfrac {L_{\rm \nu, ff}\left( \nu \right) }{h\nu }d\nu, 
\end{equation}

\noindent where $\nu_{\rm abs}$ is the frequency at which the bound-free optical depth of the CSM is 2/3 (i.e., single scatter), and depends on the structure of the CSM. In fact, $\dot Q_{\rm ion}$ is rather insensitive to $\nu_{\rm abs}$ because the frequency dependence of free-free emission is $L_{\rm \nu,ff}\propto e^{-h\nu/kT}$ (where $h$ is Planck's constant, $k$ the Boltzmann constant, and $T$ the gas temperature), and so is nearly constant up to $h\nu \approx kT$. In our simulations $kT \gtrsim 10$~$\rm keV$, and ${\nu_{\rm abs}} \lesssim 10$~$\rm keV$ even at early times, so the gamma rays being produced in the SN core easily free-stream through the CSM and do not contribute to the $\Hal$ luminosity (see also our discussion about gamma-ray free-streaming in \S~\ref{sssec:obs.brd}). To determine $L_{\rm \nu,ff}$ we use $L_{\rm \nu,ff} = 4\pi j_{\rm \nu,ff} V$, and calculate the emission $\epsilon_{\rm \nu,ff} = 4\pi j_{\rm \nu,ff}$ in each resolution element using \cite{RL1979} Equation 5.14b, where $V$ is the volume of the resolution element. We assume that $Z=7$, that at all times the shocked ejecta are fully ionized and free of hydrogen, and that the electrons are thermally distributed.

\begin{figure}
\begin{center}
\includegraphics[trim=0cm 0cm 0cm 0cm, clip=true, width=3.3in]{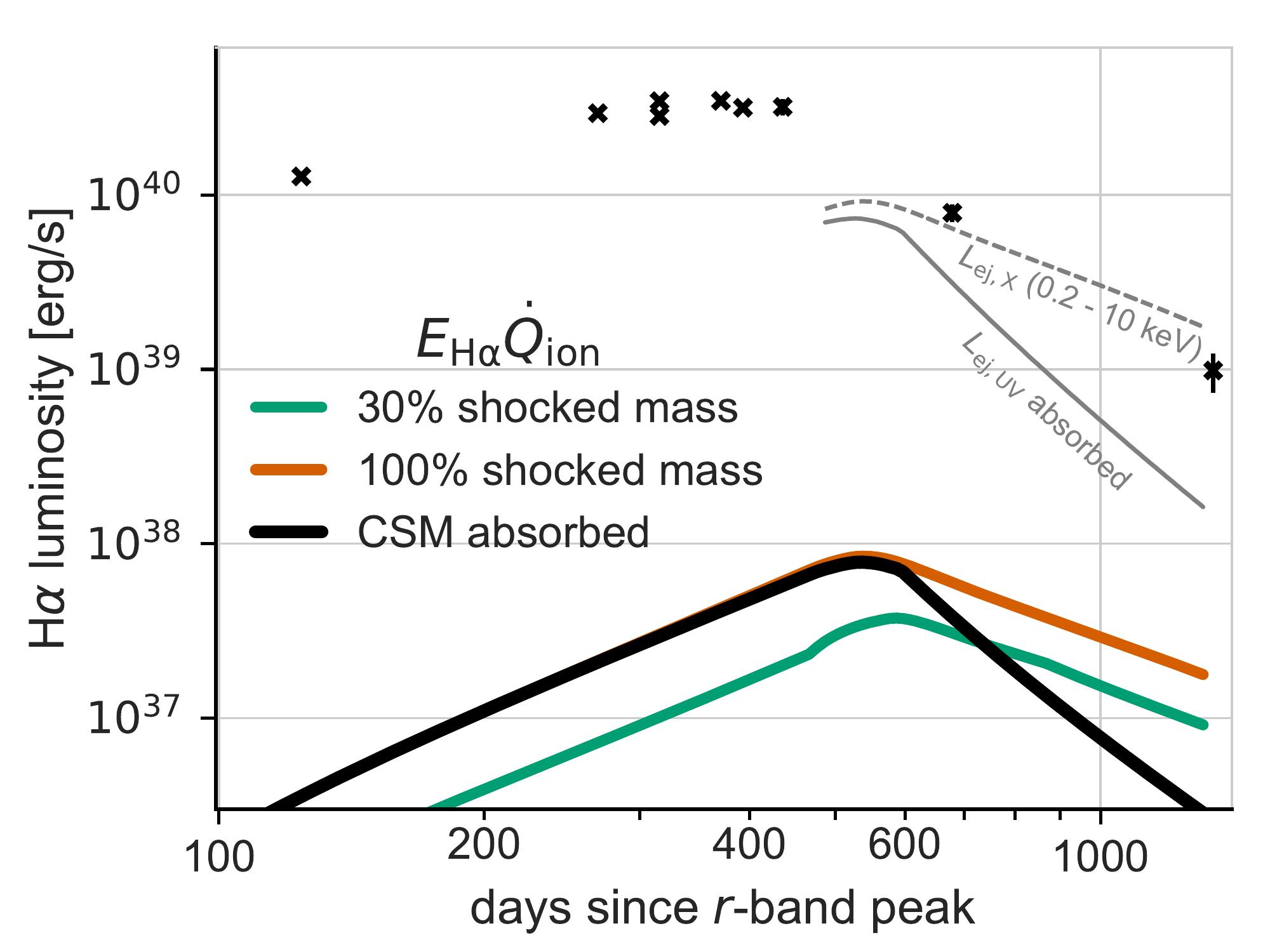}
\caption{The predicted late-time $\Hal$ luminosity from our model in which the SN ejecta impacted CSM with density $10^{-18}~{\rm g~cm^{-3}}$ and fractional width $\Delta R = 4R_{\rm imp}$ at 50 days after explosion, as described in \S~\ref{ssec:revshock}. The grey lines show the total X-ray and UV luminosity generated by the reverse shock in our model. The black curve shows the actual reprocessing in this model, which drops sharply after the CSM enters a state of free expansion due to declining optical depth to ionizing photons. For reference we show our upper limits on the reprocessed $\Hal$ luminosity: reprocessing all photons from the shocked ejecta (orange), as well as just photons produced in the inner 30\% of shocked mass (green) to account for the lack of cooling in our model. The observed late-time broad $\Hal$ luminosity is also shown for comparison (crosses). \label{fig:modelLC}}
\end{center}
\end{figure}

Based on this simulation we find that the late-time X-ray (i.e., {\it Swift} $0.2$--$10$~$\rm keV$ band) luminosity is similar to the observed $1342$ day $\Hal$ luminosity for PTF11kx (peaking at $L_{\rm X} \approx 10^{40}$ $\rm erg\ s^{-1}$ in Fig. \ref{fig:modelLC}), but with a slower evolution that is more similar to that of SN~1988Z in Figure \ref{fig:comp}. We also find that the UV luminosity absorbed by the CSM is similar to the observed $1342$ day $\Hal$ luminosity, which is below the upper limit from our {\it HST} UV imaging in \S~\ref{ssec:obs.hst}. However, we emphasize that it is inappropriate to directly compare this free-free X-ray luminosity to the observed $\Hal$ luminosity because high-energy photons must be reprocessed to produce H$\alpha$ luminosity.

We can quantitatively show that reprocessing of the X-rays by photoionization and recombination is insufficient to power the observed H$\alpha$ luminosity. As an upper limit to the possible $\Hal$ luminosity that could be generated in our model with appropriate reprocessing, we calculate $\dot{Q}_{\rm ion}$ from each mass resolution element in the shocked ejecta and convert it to $L_{\Hal}$. In Figure \ref{fig:modelLC}, this is represented by the green and orange curves. Because we do not include cooling in our hydrodynamic evolution, for reference we show the budget from just the inner 30\% of the shocked ejecta as well as from the entirety of the shocked ejecta. To actually estimate the reprocessing from the shocked CSM in our simulations requires calculating the absorbed fraction of these photons. To do this we first estimate the frequency at which the bound-free optical depth is 2/3, using the hydrogenic approximation for the bound-free cross section,

\begin{equation}\label{eq:sigmabf}
	\sigma_{\rm bf} = \frac{\pi q_e^2}{m_e c \nu_{\rm Ry}} \left( \nu/\nu_{\rm Ry} \right)^{-3},
\end{equation}

\noindent where $\nu$ is the photon frequency, $\nu_{\rm Ry} = 13.6~{\rm eV}/h$, and other constants are defined in Appendix \ref{ap:a2}. We then calculate the ionizing photon production rate up to this frequency and compare it to the hydrogen photorecombination rate of the whole CSM, which is analogous to a Str\"{o}mgren sphere calculation for ionization state, and take the minimum of the two values to be the number of recombinations occurring per second. This is shown as the bold black curve in Figure \ref{fig:modelLC}; it represents ionizing photons from all of the shocked ejecta, and we find that it falls short of the available budget once the CSM begins to freely expand, owing to the decline in optical depth. The luminosity decline is more rapid than is observed for PTF11kx, like the reverse-shock-dominated SN~1993S (see discussion in \S~\ref{sssec:obs.brd}).

Based on our analysis of this model, we conclude that although the CSM can produce X-rays with a luminosity similar to our $\Hal$ observations, photoionization and recombination in the CSM cannot explain the observed H$\alpha$, and reprocessing of the reverse shock's high-energy photons cannot reproduce the observed $\Hal$ light curve unless that reprocessing is nearly 100\% efficient. This suggests that another reprocessing mechanism must be responsible. At this point it is important to note that we have not incorporated real radiation transport in our model, or considered options such as Case C recombination, where the Balmer continuum is optically thick \citep{1992ApJ...386..181X}, or collisional excitation from the SN ejecta interacting with a multitude of low-mass CSM shells, as suggested for SNe\,Ia-CSM by \cite{2013ApJS..207....3S}. In the last two circumstances the signature observation is a large Balmer decrement (i.e., the ratio of H$\alpha$ to H$\beta$ luminosity $>10$), and S13 observed PTF11kx's Balmer decrement to increase at late times, peaking at $\sim15$ at $\sim680$ days. Unfortunately, the signal-to-noise ratio of our day $1342$ DEIMOS spectrum is too low at the blue end to derive a useful upper limit on H$\beta$, so we cannot confirm whether the Balmer decrement remains high. However, since we do find that most of the X-rays are absorbed for the conditions of interest, we think the most likely scenario is that the X-rays are thermalized and heat the CSM, leading to collisionally excited H$\alpha$ emission. 

As a final note, we consider the impact of using a CSM density in between that of D12 and our own estimate. A lower density would simply lead to a lower predicted $\Hal$ luminosity, but the higher density of D12 implied a CSM mass that was too large and caused them to infer a disk-like geometry, which is also problematic. If we raise the model's CSM density we could get an $\Hal$ emission boost, but a decrease in emitting volume (and possibly viewing-angle effects) could cause an overall reduction in the luminosity compared to spherically symmetric, lower density models. Furthermore, in a model with a $3$~M$_\odot$ CSM and $\sim1.4$~M$_\odot$ ejecta, the reverse shock will sweep over all the ejecta before the forward shock has swept over the CSM, likely rendering moot the question of reverse-shock ionization at late times. The impact of CSM characteristics like density and geometry require a more complicated modeling process than could be included in this work, but we do consider one such special case in the next section.

\subsection{The CSM as a Dense Ring}\label{ssec:disc.ring}

D12 suggested that the CSM geometry of PTF11kx was a disk instead of a shell. We consider the situation in which the CSM is a relatively dense ring, and the SN ejecta sweep over and past it, compressing and heating the material. In this situation the interaction would be continuous, and instead of being swept up the disk might survive, embedded in the ejecta, and encountering denser, lower velocity ejecta material as time passes. In the SN\,Ia ejecta, the density as a function of initial white dwarf radius ($r_{\rm WD}$) is given by
$\rho_{\rm ej} \propto r_{\rm WD}^{-10}$ if $r_{\rm WD}^{}>r_t$, and 
$\rho_{\rm ej} \propto r_{\rm WD}^{-1}$ if $r_{\rm WD}^{}<r_t$, 
where $r_t$ is a transition radius in the white dwarf, and the velocity of ejecta at the transition radius is typically $\sim5000$ $\rm km \ s^{-1}$ \citep{2010ApJ...708.1025K,2016ApJ...823..100H}. A dense ring of CSM at $R\approx 10^{16}$ $\rm cm$ would experience this transition from fast, low-density ejecta to slow, higher-density ejecta at around 230 days after explosion. This $\sim230$ day timescale matches the evolution of the broad H$\alpha$ luminosity presented by S13 (their Fig. 3), where the integrated flux increases until $>$250 days and then remains at about that level before declining. While this coincidence is interesting, and may reflect the general physical reality of timescales, distances, and velocities for the CSM-ejecta interaction, it is not enough evidence to confirm a ring geometry -- for example, a CSM distributed in dense clumps may also survive to encounter the post-transition material. Ultimately, we consider a full model of interaction between SN ejecta and alternative geometries of CSM as beyond the scope of this work.

\subsection{Interaction with the Nondegenerate Companion}\label{ssec:disc.nond}

We briefly speculate on an alternative physical interpretation for the PTF11kx observations. Like D12 and S13 before us, we have interpreted the \ion{Ca}{2} and H$\alpha$ narrow emission that arises $\sim50$ days after explosion to represent the photoionization of CSM by high-energy emission from the fastest SN\,Ia material interacting with the inner edge of this CSM. In this interpretation, the speed of the fastest SN\,Ia ejecta sets the distance between the progenitor star and the CSM to be $10^{16}$ $\rm cm$. Our alternate scenario is that the CSM is photoionized by high-energy emission from the SN\,Ia ejecta impacting the companion star. Since most binary evolution models result in a mass-donating secondary star that is very close to the white dwarf, this impact occurs almost immediately. In this scenario, the speed of light sets the distance between the progenitor star and the CSM to be $\sim 10^{17}$ $\rm cm$ (close enough to be considered CSM and not interstellar medium, $<1$~pc). The interaction of SN\,Ia ejecta with a companion star at close range is predicted to release $0.1$--$2$ $\rm keV$ X-ray emission with a total luminosity of $L_{\rm X} \approx 10^{44}$ $\rm erg\ s^{-1}$ \citep{2010ApJ...708.1025K}. CSM with a column density of $N_{\rm CSM}\approx 10^{21}$--$10^{23}$ $\rm cm^{-2}$, as we have inferred for the PTF11kx system, would be optically thick to this radiation \citep{1973AD......5...51V} and thus able to absorb and re-emit as the narrow H$\alpha$ emission observed by D12 and S13. The interaction of SN\,Ia ejecta with a companion star also predicts an optical/UV emission ``bump" on the light curve within days of explosion \citep{2010ApJ...708.1025K}, and two such detections for SNe\,Ia have been reported \citep{2015Natur.521..328C,2016ApJ...820...92M}, but PTF11kx was discovered too late to confirm or deny a similar early-time excess.

This proposed CSM at $10^{17}$ $\rm cm$ from the white dwarf would be impacted by fast ejecta with a velocity of $2\times10^4$ $\rm km\ s^{-1}$ at $\sim 580$ days after peak brightness, and indeed S13 shows an increase in the narrow H$\alpha$ line emission between observations on day $\sim430$ and 680 (see their Fig. 4). In this scenario we postulate that the appearance of broad H$\alpha$ at $40$ days after peak brightness -- previously attributed to the shocked CSM -- is instead from the stripped hydrogen-rich envelope of the companion star. The ablation and stripping of material from a star's outer layers by SN ejecta was initially investigated by \cite{1975ApJ...200..145W}. \cite{2000ApJS..128..615M} advanced this work with high-resolution simulations and found that almost the entire envelope of giant-star companions ends up embedded in the iron layers of the SN\,Ia ejecta, moving with a velocity of $v<1000$ $\rm km\ s^{-1}$. The observational signature of this is H$\alpha$ emission with FWHM $\lesssim22$ \AA, visible once the optical depth of the SN\,Ia material has dropped and photons from deep in the ejecta can escape -- typically several months after explosion. Initial searches for H$\alpha$ in the nebular-phase spectra of SNe\,Ia yielded only upper limits \citep{2007ApJ...670.1275L,2013ApJ...762L...5S,2015A&A...577A..39L}, with one recent survey reporting a tentative detection \citep{2016MNRAS.457.3254M}. Although the velocity and FWHM of PTF11kx's broad H$\alpha$ are consistent with predictions for an impacted companion (which would be on the near side because of the blueshift in peak wavelength), the relatively early appearance of this feature is not. The latter could perhaps be explained by a companion star that is significantly farther away from the primary. However, binary evolution models that result in a distant secondary also show that during the time needed for the stars to move apart, the leftover core of a post-mass-transfer red giant star has time to evolve into a white dwarf itself -- and would have no envelope to lose \citep{2012ApJ...756L...4H}. Earlier visibility of H$\alpha$ from a stripped envelope might also be attributable to differences in the radial density distribution of the ejecta, but we consider the detailed models necessary to confirm this to be beyond the scope of this work.

Ultimately, we find that detailed models are not necessary because we can identify three major detracting arguments for this speculative scenario.
(1) CSM photoionized by a single, short burst of X-ray emission would then steadily and quickly recombine, whereas Figure 1 of D12 reveals further variability (most evident in the \ion{Ca}{2}), and S13 show continued variable narrow line emission. These observations are best explained by the SN\,Ia shock proceeding through a nonuniformly distributed mass of CSM.
(2) Emission from the stripped envelope would be powered by the radioactive decay and decline accordingly, whereas we observe the broad H$\alpha$ to continue to increase in luminosity up to $\sim500$ days after peak brightness.
(3) The near-peak optical light curve of PTF11kx remained brighter for longer than most SNe\,Ia, and its early-time spectra exhibited contributions from blackbody emission that nearly overwhelmed the typical SN\,Ia features (D12, S13). If the CSM was not impacted until $>500$ days, PTF11kx would have exhibited a more typical early-time evolution. This is the strongest argument against our hypothetical scenario.

We conclude that our alternative model is not supported by the observations, but it was worthwhile to consider given the ongoing attempts to find signatures of a nondegenerate companion's stripped material in SNe\,Ia. We concede that it remains difficult to distinguish an H-rich CSM from an H-rich companion (see, e.g., \citealt{2016ApJ...820...92M} vs. \citealt{2016arXiv161007601S}); as mentioned above, early light-curve observations help to constrain direct interaction between the ejecta and a nearby companion star. In the future, being able to differentiate these effects will lead to a better understanding of the late stages of evolution of SN\,Ia progenitor systems. 

\section{Conclusions}\label{sec:con}

We present spectroscopic observations of the SN\,Ia-CSM PTF11kx at 3.5 years post-explosion, showing that the broad component of H$\alpha$ emission has continued the decline in luminosity that was first noted after $\sim500$ days. S13 suggested that this decline marked the time at which the SN ejecta overtake the majority of the CSM, and we establish that this is consistent with our observed lack of narrow H$\alpha$ emission from PTF11kx at late times -- there is no longer much (or any) unshocked material beyond the ejecta front to be ionized by the interaction -- and also with the late-time H$\alpha$ luminosity decline for other CSM-interaction dominated SNe. We demonstrate that at 1342 days, the broad H$\alpha$ appears to have the same blueshifted velocity and FWHM as it did at $\sim100$ days after peak brightness (D12; S13). If this blueshift is a result of the red side of the line being preferentially absorbed by newly formed dust, then the dust has not yet been destroyed by the reverse shock (however, it would still be early for that to happen because the reverse shock propagates more slowly through the remnant). We show that the presence of newly formed dust is also indicated by our late-time IR imaging observations from {\it Spitzer}, which furthermore place the dust at the same radius as the shocked CSM and suggest that it may be collisionally heated.

We discuss our latest observations in context with the previously reported data for PTF11kx at early and late times, in some cases re-interpreting the early-time spectra, and revise the previously published physical parameters of the CSM: column density, radial extent, and total mass. We show that the radial extent of the CSM is likely to be thick ($\Delta R_{\rm CSM} \approx 4 R_{\rm CSM}$), and that the column and volume particle densities (and, consequently, the total mass) may be significantly lower ($M_{\rm CSM}\approx 0.06$ M$_{\odot}$) than previously reported. We find that this CSM mass is similar to that of other SN\,Ia-CSM events, but larger than that predicted by models of recurrent novae, and that it indicates a low gas-to-dust ratio in the dust-forming region for PTF11kx. We create a model of the CSM interaction to demonstrate that sufficient amounts of high-energy photons exist to power the late-time H$\alpha$ emission, but that reprocessing by photoionization and recombination are insufficient, leaving thermalization and collisional excitation as the most likely physical explanation for the observed H$\alpha$ luminosity at late times. We briefly consider a ring geometry for the CSM but find it difficult to confirm an alternative geometry with the minimal data in hand. Finally, we show how an alternative physical interpretation of SN ejecta impacting a nondegenerate companion star instead of CSM created by it can be rejected for PTF11kx.

At the time of publication, six years after its explosion, PTF11kx remains unique in terms of its unambiguous SN spectral classification as a Type Ia and clear signatures of ejecta interaction with a hydrogren-rich CSM. Since wide-field SN surveys have only grown larger, and spectroscopic follow-up observations more automated, it would appear that such events are intrinsically very rare. To constrain the true occurrence rate of PTF11kx-like SNe\,Ia, and thereby the fraction with nondegenerate companion stars, we may have to wait until the Large Synoptic Survey Telescope begins full operations in 2022 \citep{2008arXiv0805.2366I}. During this time the SN\,Ia-CSM candidates will have to be photometrically identified in order to optimize the relatively limited amount of spectroscopic follow-up time that will be available to pursue this science goal, and we hope that this work will motivate and guide those future endeavors.

\section*{Acknowledgements}

This work is based on observations from the Low Resolution Imaging Spectrometer at the Keck-1 telescope and the DEep Imaging Multi-Object Spectrograph at the Keck-2 telescope. We are grateful to the staff at Keck Observatory for their assistance. The W.~M.\ Keck Observatory is operated as a scientific partnership among the California Institute of Technology, the University of California, and the National Aeronautics and Space Administration (NASA); it was made possible by the generous financial support of the W.~M.\ Keck Foundation. We extend special thanks to those of Hawaiian ancestry on whose sacred mountain we are privileged to be guests. We thank Daniel Perley and Brad Cenko for the use of, and assistance with, their Keck LRIS imaging and spectroscopy reduction pipeline\footnote{Dan Perley's pipeline can be found at \url{http://www.astro.caltech.edu/$\sim$dperley/programs/lpipe.html}}.

This work is based in part on observations made with the {\it Spitzer Space Telescope}, which is operated by the Jet Propulsion Laboratory, California Institute of Technology, under a contract with NASA. It is also based in part on observations made with the NASA/ESA {\it Hubble Space Telescope}, obtained [from the Data Archive] at the Space Telescope Science Institute (STScI), which is operated by the Association of Universities for Research in Astronomy (AURA), Inc., under NASA contract NAS 5-26555. The new observations are associated with program GO-14779. This research has made use of the NASA/IPAC Extragalactic Database (NED), which is operated by the Jet Propulsion Laboratory, California Institute of Technology, under contract with NASA.

Support for {\it HST} programs GO-14779 and AR-14295 was provided by NASA through grants from STScI, which is operated by AURA, Inc., under NASA contract NAS5-26555. The supernova research of A.V.F.'s group at U.C. Berkeley is supported by Gary \& Cynthia Bengier, the Richard \& Rhoda Goldman Fund, the Christopher R. Redlich Fund, the TABASGO Foundation, and National Science Foundation (NSF) grant AST--1211916. J.M.S. was supported by an NSF Astronomy and Astrophysics Postdoctoral Fellowship under award AST--1302771.

\bibliographystyle{apj}
\bibliography{apj-jour,myrefs}

\begin{thebibliography}{}
\expandafter\ifx\csname natexlab\endcsname\relax\def\natexlab#1{#1}\fi

\bibitem[{{Aihara} {et~al.}(2011){Aihara}, {Allende Prieto}, {An}, {Anderson},
  {Aubourg}, {Balbinot}, {Beers}, {Berlind}, {Bickerton}, {Bizyaev}, {Blanton},
  {Bochanski}, {Bolton}, {Bovy}, {Brandt}, {Brinkmann}, {Brown}, {Brownstein},
  {Busca}, {Campbell}, {Carr}, {Chen}, {Chiappini}, {Comparat}, {Connolly},
  {Cortes}, {Croft}, {Cuesta}, {da Costa}, {Davenport}, {Dawson}, {Dhital},
  {Ealet}, {Ebelke}, {Edmondson}, {Eisenstein}, {Escoffier}, {Esposito},
  {Evans}, {Fan}, {Femen{\'{\i}}a Castell{\'a}}, {Font-Ribera}, {Frinchaboy},
  {Ge}, {Gillespie}, {Gilmore}, {Gonz{\'a}lez Hern{\'a}ndez}, {Gott}, {Gould},
  {Grebel}, {Gunn}, {Hamilton}, {Harding}, {Harris}, {Hawley}, {Hearty}, {Ho},
  {Hogg}, {Holtzman}, {Honscheid}, {Inada}, {Ivans}, {Jiang}, {Johnson},
  {Jordan}, {Jordan}, {Kazin}, {Kirkby}, {Klaene}, {Knapp}, {Kneib},
  {Kochanek}, {Koesterke}, {Kollmeier}, {Kron}, {Lampeitl}, {Lang}, {Le Goff},
  {Lee}, {Lin}, {Long}, {Loomis}, {Lucatello}, {Lundgren}, {Lupton}, {Ma},
  {MacDonald}, {Mahadevan}, {Maia}, {Makler}, {Malanushenko}, {Malanushenko},
  {Mandelbaum}, {Maraston}, {Margala}, {Masters}, {McBride}, {McGehee},
  {McGreer}, {M{\'e}nard}, {Miralda-Escud{\'e}}, {Morrison}, {Mullally},
  {Muna}, {Munn}, {Murayama}, {Myers}, {Naugle}, {Neto}, {Nguyen}, {Nichol},
  {O'Connell}, {Ogando}, {Olmstead}, {Oravetz}, {Padmanabhan},
  {Palanque-Delabrouille}, {Pan}, {Pandey}, {P{\^a}ris}, {Percival},
  {Petitjean}, {Pfaffenberger}, {Pforr}, {Phleps}, {Pichon}, {Pieri}, {Prada},
  {Price-Whelan}, {Raddick}, {Ramos}, {Reyl{\'e}}, {Rich}, {Richards}, {Rix},
  {Robin}, {Rocha-Pinto}, {Rockosi}, {Roe}, {Rollinde}, {Ross}, {Ross},
  {Rossetto}, {S{\'a}nchez}, {Sayres}, {Schlegel}, {Schlesinger}, {Schmidt},
  {Schneider}, {Sheldon}, {Shu}, {Simmerer}, {Simmons}, {Sivarani}, {Snedden},
  {Sobeck}, {Steinmetz}, {Strauss}, {Szalay}, {Tanaka}, {Thakar}, {Thomas},
  {Tinker}, {Tofflemire}, {Tojeiro}, {Tremonti}, {Vandenberg}, {Vargas
  Maga{\~n}a}, {Verde}, {Vogt}, {Wake}, {Wang}, {Weaver}, {Weinberg}, {White},
  {White}, {Yanny}, {Yasuda}, {Yeche}, \& {Zehavi}}]{2011ApJS..193...29A}
{Aihara}, H., {Allende Prieto}, C., {An}, D., {et~al.} 2011, \apjs, 193, 29

\bibitem[{{Aretxaga} {et~al.}(1999){Aretxaga}, {Benetti}, {Terlevich},
  {Fabian}, {Cappellaro}, {Turatto}, \& {della Valle}}]{1999MNRAS.309..343A}
{Aretxaga}, I., {Benetti}, S., {Terlevich}, R.~J., {et~al.} 1999, \mnras, 309,
  343

\bibitem[{{Bianco} {et~al.}(2011){Bianco}, {Howell}, {Sullivan}, {Conley},
  {Kasen}, {Gonz{\'a}lez-Gait{\'a}n}, {Guy}, {Astier}, {Balland}, {Carlberg},
  {Fouchez}, {Fourmanoit}, {Hardin}, {Hook}, {Lidman}, {Pain},
  {Palanque-Delabrouille}, {Perlmutter}, {Perrett}, {Pritchet}, {Regnault},
  {Rich}, \& {Ruhlmann-Kleider}}]{2011ApJ...741...20B}
{Bianco}, F.~B., {Howell}, D.~A., {Sullivan}, M., {et~al.} 2011, \apj, 741, 20

\bibitem[{{Bloom} {et~al.}(2012){Bloom}, {Kasen}, {Shen}, {Nugent}, {Butler},
  {Graham}, {Howell}, {Kolb}, {Holmes}, {Haswell}, {Burwitz}, {Rodriguez}, \&
  {Sullivan}}]{2012ApJ...744L..17B}
{Bloom}, J.~S., {Kasen}, D., {Shen}, K.~J., {et~al.} 2012, \apjl, 744, L17

\bibitem[{{Cao} {et~al.}(2015){Cao}, {Kulkarni}, {Howell}, {Gal-Yam},
  {Kasliwal}, {Valenti}, {Johansson}, {Amanullah}, {Goobar}, {Sollerman},
  {Taddia}, {Horesh}, {Sagiv}, {Cenko}, {Nugent}, {Arcavi}, {Surace},
  {Wo{\'z}niak}, {Moody}, {Rebbapragada}, {Bue}, \&
  {Gehrels}}]{2015Natur.521..328C}
{Cao}, Y., {Kulkarni}, S.~R., {Howell}, D.~A., {et~al.} 2015, \nat, 521, 328

\bibitem[{{Chevalier} \& {Fransson}(1994)}]{1994ApJ...420..268C}
{Chevalier}, R.~A., \& {Fransson}, C. 1994, \apj, 420, 268

\bibitem[{{Dilday} {et~al.}(2012){Dilday}, {Howell}, {Cenko}, {Silverman},
  {Nugent}, {Sullivan}, {Ben-Ami}, {Bildsten}, {Bolte}, {Endl}, {Filippenko},
  {Gnat}, {Horesh}, {Hsiao}, {Kasliwal}, {Kirkman}, {Maguire}, {Marcy},
  {Moore}, {Pan}, {Parrent}, {Podsiadlowski}, {Quimby}, {Sternberg}, {Suzuki},
  {Tytler}, {Xu}, {Bloom}, {Gal-Yam}, {Hook}, {Kulkarni}, {Law}, {Ofek},
  {Polishook}, \& {Poznanski}}]{2012Sci...337..942D}
{Dilday}, B., {Howell}, D.~A., {Cenko}, S.~B., {et~al.} 2012, Science, 337, 942

\bibitem[{{Faber} {et~al.}(2003){Faber}, {Phillips}, {Kibrick}, {Alcott},
  {Allen}, {Burrous}, {Cantrall}, {Clarke}, {Coil}, {Cowley}, {Davis}, {Deich},
  {Dietsch}, {Gilmore}, {Harper}, {Hilyard}, {Lewis}, {McVeigh}, {Newman},
  {Osborne}, {Schiavon}, {Stover}, {Tucker}, {Wallace}, {Wei}, {Wirth}, \&
  {Wright}}]{2003SPIE.4841.1657F}
{Faber}, S.~M., {Phillips}, A.~C., {Kibrick}, R.~I., {et~al.} 2003, in Society
  of Photo-Optical Instrumentation Engineers (SPIE) Conference Series, Vol.
  4841, Instrument Design and Performance for Optical/Infrared Ground-based
  Telescopes, ed. M.~{Iye} \& A.~F.~M. {Moorwood}, 1657--1669

\bibitem[{{Fassia} {et~al.}(2001){Fassia}, {Meikle}, {Chugai}, {Geballe},
  {Lundqvist}, {Walton}, {Pollacco}, {Veilleux}, {Wright}, {Pettini}, {Kerr},
  {Puchnarewicz}, {Puxley}, {Irwin}, {Packham}, {Smartt}, \&
  {Harmer}}]{2001MNRAS.325..907F}
{Fassia}, A., {Meikle}, W.~P.~S., {Chugai}, N., {et~al.} 2001, \mnras, 325, 907

\bibitem[{{Fazio} {et~al.}(2004){Fazio}, {Hora}, {Allen}, {Ashby}, {Barmby},
  {Deutsch}, {Huang}, {Kleiner}, {Marengo}, {Megeath}, {Melnick}, {Pahre},
  {Patten}, {Polizotti}, {Smith}, {Taylor}, {Wang}, {Willner}, {Hoffmann},
  {Pipher}, {Forrest}, {McMurty}, {McCreight}, {McKelvey}, {McMurray}, {Koch},
  {Moseley}, {Arendt}, {Mentzell}, {Marx}, {Losch}, {Mayman}, {Eichhorn},
  {Krebs}, {Jhabvala}, {Gezari}, {Fixsen}, {Flores}, {Shakoorzadeh}, {Jungo},
  {Hakun}, {Workman}, {Karpati}, {Kichak}, {Whitley}, {Mann}, {Tollestrup},
  {Eisenhardt}, {Stern}, {Gorjian}, {Bhattacharya}, {Carey}, {Nelson},
  {Glaccum}, {Lacy}, {Lowrance}, {Laine}, {Reach}, {Stauffer}, {Surace},
  {Wilson}, {Wright}, {Hoffman}, {Domingo}, \& {Cohen}}]{2004ApJS..154...10F}
{Fazio}, G.~G., {Hora}, J.~L., {Allen}, L.~E., {et~al.} 2004, \apjs, 154, 10

\bibitem[{{Filippenko} {et~al.}(1992){Filippenko}, {Richmond}, {Matheson},
  {Shields}, {Burbidge}, {Cohen}, {Dickinson}, {Malkan}, {Nelson}, {Pietz},
  {Schlegel}, {Schmeer}, {Spinrad}, {Steidel}, {Tran}, \&
  {Wren}}]{1992ApJ...384L..15F}
{Filippenko}, A.~V., {Richmond}, M.~W., {Matheson}, T., {et~al.} 1992, \apjl,
  384, L15

\bibitem[{{Fox} {et~al.}(2010){Fox}, {Chevalier}, {Dwek}, {Skrutskie},
  {Sugerman}, \& {Leisenring}}]{2010ApJ...725.1768F}
{Fox}, O.~D., {Chevalier}, R.~A., {Dwek}, E., {et~al.} 2010, \apj, 725, 1768

\bibitem[{{Fox} \& {Filippenko}(2013)}]{2013ApJ...772L...6F}
{Fox}, O.~D., \& {Filippenko}, A.~V. 2013, \apjl, 772, L6

\bibitem[{{Fox} {et~al.}(2013){Fox}, {Filippenko}, {Skrutskie}, {Silverman},
  {Ganeshalingam}, {Cenko}, \& {Clubb}}]{2013AJ....146....2F}
{Fox}, O.~D., {Filippenko}, A.~V., {Skrutskie}, M.~F., {et~al.} 2013, \aj, 146,
  2

\bibitem[{{Fox} {et~al.}(2015){Fox}, {Silverman}, {Filippenko}, {Mauerhan},
  {Becker}, {Borish}, {Cenko}, {Clubb}, {Graham}, {Hsiao}, {Kelly}, {Lee},
  {Marion}, {Milisavljevic}, {Parrent}, {Shivvers}, {Skrutskie}, {Smith},
  {Wilson}, \& {Zheng}}]{2015MNRAS.447..772F}
{Fox}, O.~D., {Silverman}, J.~M., {Filippenko}, A.~V., {et~al.} 2015, \mnras,
  447, 772

\bibitem[{{Fox} {et~al.}(2016){Fox}, {Johansson}, {Kasliwal}, {Andrews},
  {Bally}, {Bond}, {Boyer}, {Gehrz}, {Helou}, {Hsiao}, {Masci},
  {Parthasarathy}, {Smith}, {Tinyanont}, \& {Van Dyk}}]{2016ApJ...816L..13F}
{Fox}, O.~D., {Johansson}, J., {Kasliwal}, M., {et~al.} 2016, \apjl, 816, L13

\bibitem[{{Gerardy} {et~al.}(2004){Gerardy}, {H{\"o}flich}, {Fesen}, {Marion},
  {Nomoto}, {Quimby}, {Schaefer}, {Wang}, \& {Wheeler}}]{2004ApJ...607..391G}
{Gerardy}, C.~L., {H{\"o}flich}, P., {Fesen}, R.~A., {et~al.} 2004, \apj, 607,
  391

\bibitem[{{Graham} {et~al.}(2015){Graham}, {Nugent}, {Sullivan}, {Filippenko},
  {Cenko}, {Silverman}, {Clubb}, \& {Zheng}}]{2015MNRAS.454.1948G}
{Graham}, M.~L., {Nugent}, P.~E., {Sullivan}, M., {et~al.} 2015, \mnras, 454,
  1948

\bibitem[{{Graur} {et~al.}(2014){Graur}, {Maoz}, \&
  {Shara}}]{2014MNRAS.442L..28G}
{Graur}, O., {Maoz}, D., \& {Shara}, M.~M. 2014, \mnras, 442, L28

\bibitem[{{Graur} {et~al.}(2016){Graur}, {Zurek}, {Shara}, {Riess},
  {Seitenzahl}, \& {Rest}}]{2016ApJ...819...31G}
{Graur}, O., {Zurek}, D., {Shara}, M.~M., {et~al.} 2016, \apj, 819, 31

\bibitem[{{Hachisu} {et~al.}(2012){Hachisu}, {Kato}, \&
  {Nomoto}}]{2012ApJ...756L...4H}
{Hachisu}, I., {Kato}, M., \& {Nomoto}, K. 2012, \apjl, 756, L4

\bibitem[{{Harris} {et~al.}(2016){Harris}, {Nugent}, \&
  {Kasen}}]{2016ApJ...823..100H}
{Harris}, C.~E., {Nugent}, P.~E., \& {Kasen}, D.~N. 2016, \apj, 823, 100

\bibitem[{{Howell}(2011)}]{2011NatCo...2E.350H}
{Howell}, D.~A. 2011, Nature Communications, 2, 350

\bibitem[{{Ivezi{\'c}} {et~al.}(2008){Ivezi{\'c}}, {Tyson}, {Abel}, {Acosta},
  {Allsman}, {AlSayyad}, {Anderson}, {Andrew}, {Angel}, {Angeli}, {Ansari},
  {Antilogus}, {Arndt}, {Astier}, {Aubourg}, {Axelrod}, {Bard}, {Barr},
  {Barrau}, {Bartlett}, {Bauman}, {Beaumont}, {Becker}, {Becla}, {Beldica},
  {Bellavia}, {Blanc}, {Blandford}, {Bloom}, {Bogart}, {Borne}, {Bosch},
  {Boutigny}, {Brandt}, {Brown}, {Bullock}, {Burchat}, {Burke}, {Cagnoli},
  {Calabrese}, {Chandrasekharan}, {Chesley}, {Cheu}, {Chiang}, {Claver},
  {Connolly}, {Cook}, {Cooray}, {Covey}, {Cribbs}, {Cui}, {Cutri}, {Daubard},
  {Daues}, {Delgado}, {Digel}, {Doherty}, {Dubois}, {Dubois-Felsmann},
  {Durech}, {Eracleous}, {Ferguson}, {Frank}, {Freemon}, {Gangler}, {Gawiser},
  {Geary}, {Gee}, {Geha}, {Gibson}, {Gilmore}, {Glanzman}, {Goodenow},
  {Gressler}, {Gris}, {Guyonnet}, {Hascall}, {Haupt}, {Hernandez}, {Hogan},
  {Huang}, {Huffer}, {Innes}, {Jacoby}, {Jain}, {Jee}, {Jernigan},
  {Jevremovic}, {Johns}, {Jones}, {Juramy-Gilles}, {Juri{\'c}}, {Kahn},
  {Kalirai}, {Kallivayalil}, {Kalmbach}, {Kantor}, {Kasliwal}, {Kessler},
  {Kirkby}, {Knox}, {Kotov}, {Krabbendam}, {Krughoff}, {Kubanek}, {Kuczewski},
  {Kulkarni}, {Lambert}, {Le Guillou}, {Levine}, {Liang}, {Lim}, {Lintott},
  {Lupton}, {Mahabal}, {Marshall}, {Marshall}, {May}, {McKercher}, {Migliore},
  {Miller}, {Mills}, {Monet}, {Moniez}, {Neill}, {Nief}, {Nomerotski},
  {Nordby}, {O'Connor}, {Oliver}, {Olivier}, {Olsen}, {Ortiz}, {Owen}, {Pain},
  {Peterson}, {Petry}, {Pierfederici}, {Pietrowicz}, {Pike}, {Pinto}, {Plante},
  {Plate}, {Price}, {Prouza}, {Radeka}, {Rajagopal}, {Rasmussen}, {Regnault},
  {Ridgway}, {Ritz}, {Rosing}, {Roucelle}, {Rumore}, {Russo}, {Saha},
  {Sassolas}, {Schalk}, {Schindler}, {Schneider}, {Schumacher}, {Sebag},
  {Sembroski}, {Seppala}, {Shipsey}, {Silvestri}, {Smith}, {Smith}, {Strauss},
  {Stubbs}, {Sweeney}, {Szalay}, {Takacs}, {Thaler}, {Van Berg}, {Vanden Berk},
  {Vetter}, {Virieux}, {Xin}, {Walkowicz}, {Walter}, {Wang}, {Warner},
  {Willman}, {Wittman}, {Wolff}, {Wood-Vasey}, {Yoachim}, {Zhan}, \& {for the
  LSST Collaboration}}]{2008arXiv0805.2366I}
{Ivezi{\'c}}, Z., {Tyson}, J.~A., {Abel}, B., {et~al.} 2008, ArXiv e-prints,
  arXiv:0805.2366

\bibitem[{{Kasen}(2010)}]{2010ApJ...708.1025K}
{Kasen}, D. 2010, \apj, 708, 1025

\bibitem[{{Katsuda} {et~al.}(2015){Katsuda}, {Mori}, {Maeda}, {Tanaka},
  {Koyama}, {Tsunemi}, {Nakajima}, {Maeda}, {Ozaki}, \&
  {Petre}}]{2015ApJ...808...49K}
{Katsuda}, S., {Mori}, K., {Maeda}, K., {et~al.} 2015, \apj, 808, 49

\bibitem[{{Law} {et~al.}(2009){Law}, {Kulkarni}, {Dekany}, {Ofek}, {Quimby},
  {Nugent}, {Surace}, {Grillmair}, {Bloom}, {Kasliwal}, {Bildsten}, {Brown},
  {Cenko}, {Ciardi}, {Croner}, {Djorgovski}, {van Eyken}, {Filippenko}, {Fox},
  {Gal-Yam}, {Hale}, {Hamam}, {Helou}, {Henning}, {Howell}, {Jacobsen},
  {Laher}, {Mattingly}, {McKenna}, {Pickles}, {Poznanski}, {Rahmer}, {Rau},
  {Rosing}, {Shara}, {Smith}, {Starr}, {Sullivan}, {Velur}, {Walters}, \&
  {Zolkower}}]{2009PASP..121.1395L}
{Law}, N.~M., {Kulkarni}, S.~R., {Dekany}, R.~G., {et~al.} 2009, \pasp, 121,
  1395

\bibitem[{{Leloudas} {et~al.}(2015){Leloudas}, {Hsiao}, {Johansson}, {Maeda},
  {Moriya}, {Nordin}, {Petrushevska}, {Silverman}, {Sollerman}, {Stritzinger},
  {Taddia}, \& {Xu}}]{2015A&A...574A..61L}
{Leloudas}, G., {Hsiao}, E.~Y., {Johansson}, J., {et~al.} 2015, \aap, 574, A61

\bibitem[{{Leonard}(2007)}]{2007ApJ...670.1275L}
{Leonard}, D.~C. 2007, \apj, 670, 1275

\bibitem[{{Li} {et~al.}(2011){Li}, {Bloom}, {Podsiadlowski}, {Miller}, {Cenko},
  {Jha}, {Sullivan}, {Howell}, {Nugent}, {Butler}, {Ofek}, {Kasliwal},
  {Richards}, {Stockton}, {Shih}, {Bildsten}, {Shara}, {Bibby}, {Filippenko},
  {Ganeshalingam}, {Silverman}, {Kulkarni}, {Law}, {Poznanski}, {Quimby},
  {McCully}, {Patel}, {Maguire}, \& {Shen}}]{2011Natur.480..348L}
{Li}, W., {Bloom}, J.~S., {Podsiadlowski}, P., {et~al.} 2011, \nat, 480, 348

\bibitem[{{Lundqvist} {et~al.}(2015){Lundqvist}, {Nyholm}, {Taddia},
  {Sollerman}, {Johansson}, {Kozma}, {Lundqvist}, {Fransson}, {Garnavich},
  {Kromer}, {Shappee}, \& {Goobar}}]{2015A&A...577A..39L}
{Lundqvist}, P., {Nyholm}, A., {Taddia}, F., {et~al.} 2015, \aap, 577, A39

\bibitem[{{Maguire} {et~al.}(2016){Maguire}, {Taubenberger}, {Sullivan}, \&
  {Mazzali}}]{2016MNRAS.457.3254M}
{Maguire}, K., {Taubenberger}, S., {Sullivan}, M., \& {Mazzali}, P.~A. 2016,
  \mnras, 457, 3254

\bibitem[{{Marietta} {et~al.}(2000){Marietta}, {Burrows}, \&
  {Fryxell}}]{2000ApJS..128..615M}
{Marietta}, E., {Burrows}, A., \& {Fryxell}, B. 2000, \apjs, 128, 615

\bibitem[{{Marion} {et~al.}(2016){Marion}, {Brown}, {Vink{\'o}}, {Silverman},
  {Sand}, {Challis}, {Kirshner}, {Wheeler}, {Berlind}, {Brown}, {Calkins},
  {Camacho}, {Dhungana}, {Foley}, {Friedman}, {Graham}, {Howell}, {Hsiao},
  {Irwin}, {Jha}, {Kehoe}, {Macri}, {Maeda}, {Mandel}, {McCully}, {Pandya},
  {Rines}, {Wilhelmy}, \& {Zheng}}]{2016ApJ...820...92M}
{Marion}, G.~H., {Brown}, P.~J., {Vink{\'o}}, J., {et~al.} 2016, \apj, 820, 92

\bibitem[{{Matsuura} {et~al.}(2015){Matsuura}, {Dwek}, {Barlow}, {Babler},
  {Baes}, {Meixner}, {Cernicharo}, {Clayton}, {Dunne}, {Fransson}, {Fritz},
  {Gear}, {Gomez}, {Groenewegen}, {Indebetouw}, {Ivison}, {Jerkstrand},
  {Lebouteiller}, {Lim}, {Lundqvist}, {Pearson}, {Roman-Duval}, {Royer},
  {Staveley-Smith}, {Swinyard}, {van Hoof}, {van Loon}, {Verstappen}, {Wesson},
  {Zanardo}, {Blommaert}, {Decin}, {Reach}, {Sonneborn}, {Van de Steene}, \&
  {Yates}}]{2015ApJ...800...50M}
{Matsuura}, M., {Dwek}, E., {Barlow}, M.~J., {et~al.} 2015, \apj, 800, 50

\bibitem[{{Mauerhan} \& {Smith}(2012)}]{2012MNRAS.424.2659M}
{Mauerhan}, J., \& {Smith}, N. 2012, \mnras, 424, 2659

\bibitem[{{Moore} \& {Bildsten}(2012)}]{2012ApJ...761..182M}
{Moore}, K., \& {Bildsten}, L. 2012, \apj, 761, 182

\bibitem[{{Noda} {et~al.}(2016){Noda}, {Suda}, \&
  {Shigeyama}}]{2016PASJ...68...11N}
{Noda}, K., {Suda}, T., \& {Shigeyama}, T. 2016, \pasj, 68, 11

\bibitem[{{Nugent} {et~al.}(2011){Nugent}, {Sullivan}, {Cenko}, {Thomas},
  {Kasen}, {Howell}, {Bersier}, {Bloom}, {Kulkarni}, {Kandrashoff},
  {Filippenko}, {Silverman}, {Marcy}, {Howard}, {Isaacson}, {Maguire},
  {Suzuki}, {Tarlton}, {Pan}, {Bildsten}, {Fulton}, {Parrent}, {Sand},
  {Podsiadlowski}, {Bianco}, {Dilday}, {Graham}, {Lyman}, {James}, {Kasliwal},
  {Law}, {Quimby}, {Hook}, {Walker}, {Mazzali}, {Pian}, {Ofek}, {Gal-Yam}, \&
  {Poznanski}}]{2011Natur.480..344N}
{Nugent}, P.~E., {Sullivan}, M., {Cenko}, S.~B., {et~al.} 2011, \nat, 480, 344

\bibitem[{{Oke} {et~al.}(1995){Oke}, {Cohen}, {Carr}, {Cromer}, {Dingizian},
  {Harris}, {Labrecque}, {Lucinio}, {Schaal}, {Epps}, \&
  {Miller}}]{1995PASP..107..375O}
{Oke}, J.~B., {Cohen}, J.~G., {Carr}, M., {et~al.} 1995, \pasp, 107, 375

\bibitem[{{Pan} {et~al.}(2012){Pan}, {Ricker}, \& {Taam}}]{2012ApJ...760...21P}
{Pan}, K.-C., {Ricker}, P.~M., \& {Taam}, R.~E. 2012, \apj, 760, 21

\bibitem[{{Pan} {et~al.}(2013){Pan}, {Ricker}, \& {Taam}}]{2013ApJ...773...49P}
---. 2013, \apj, 773, 49

\bibitem[{{Patat} {et~al.}(2007){Patat}, {Chandra}, {Chevalier}, {Justham},
  {Podsiadlowski}, {Wolf}, {Gal-Yam}, {Pasquini}, {Crawford}, {Mazzali},
  {Pauldrach}, {Nomoto}, {Benetti}, {Cappellaro}, {Elias-Rosa}, {Hillebrandt},
  {Leonard}, {Pastorello}, {Renzini}, {Sabbadin}, {Simon}, \&
  {Turatto}}]{2007Sci...317..924P}
{Patat}, F., {Chandra}, P., {Chevalier}, R., {et~al.} 2007, Science, 317, 924

\bibitem[{{Podsiadlowski}(2003)}]{2003astro.ph..3660P}
{Podsiadlowski}, P. 2003, ArXiv Astrophysics e-prints, astro-ph/0303660

\bibitem[{{Pozzo} {et~al.}(2004){Pozzo}, {Meikle}, {Fassia}, {Geballe},
  {Lundqvist}, {Chugai}, \& {Sollerman}}]{2004MNRAS.352..457P}
{Pozzo}, M., {Meikle}, W.~P.~S., {Fassia}, A., {et~al.} 2004, \mnras, 352, 457

\bibitem[{{Rau} {et~al.}(2009){Rau}, {Kulkarni}, {Law}, {Bloom}, {Ciardi},
  {Djorgovski}, {Fox}, {Gal-Yam}, {Grillmair}, {Kasliwal}, {Nugent}, {Ofek},
  {Quimby}, {Reach}, {Shara}, {Bildsten}, {Cenko}, {Drake}, {Filippenko},
  {Helfand}, {Helou}, {Howell}, {Poznanski}, \&
  {Sullivan}}]{2009PASP..121.1334R}
{Rau}, A., {Kulkarni}, S.~R., {Law}, N.~M., {et~al.} 2009, \pasp, 121, 1334

\bibitem[{{Rybicki} \& {Lightman}(1979)}]{RL1979}
{Rybicki}, G.~B., \& {Lightman}, A.~P. 1979, {Radiative processes in
  astrophysics} (John Wiley \& Sons, Inc.)

\bibitem[{{Schlegel} {et~al.}(1998){Schlegel}, {Finkbeiner}, \&
  {Davis}}]{1998ApJ...500..525S}
{Schlegel}, D.~J., {Finkbeiner}, D.~P., \& {Davis}, M. 1998, \apj, 500, 525

\bibitem[{{Shappee} {et~al.}(2016){Shappee}, {Piro}, {Stanek}, {Patel},
  {Margutti}, {Lipunov}, \& {Pogge}}]{2016arXiv161007601S}
{Shappee}, B.~J., {Piro}, A.~L., {Stanek}, K.~Z., {et~al.} 2016, ArXiv
  e-prints, arXiv:1610.07601

\bibitem[{{Shappee} {et~al.}(2013){Shappee}, {Stanek}, {Pogge}, \&
  {Garnavich}}]{2013ApJ...762L...5S}
{Shappee}, B.~J., {Stanek}, K.~Z., {Pogge}, R.~W., \& {Garnavich}, P.~M. 2013,
  \apjl, 762, L5

\bibitem[{{Silverman} {et~al.}(2013{\natexlab{a}}){Silverman}, {Nugent},
  {Gal-Yam}, {Sullivan}, {Howell}, {Filippenko}, {Pan}, {Cenko}, \&
  {Hook}}]{2013ApJ...772..125S}
{Silverman}, J.~M., {Nugent}, P.~E., {Gal-Yam}, A., {et~al.}
  2013{\natexlab{a}}, \apj, 772, 125

\bibitem[{{Silverman} {et~al.}(2013{\natexlab{b}}){Silverman}, {Nugent},
  {Gal-Yam}, {Sullivan}, {Howell}, {Filippenko}, {Arcavi}, {Ben-Ami}, {Bloom},
  {Cenko}, {Cao}, {Chornock}, {Clubb}, {Coil}, {Foley}, {Graham}, {Griffith},
  {Horesh}, {Kasliwal}, {Kulkarni}, {Leonard}, {Li}, {Matheson}, {Miller},
  {Modjaz}, {Ofek}, {Pan}, {Perley}, {Poznanski}, {Quimby}, {Steele},
  {Sternberg}, {Xu}, \& {Yaron}}]{2013ApJS..207....3S}
---. 2013{\natexlab{b}}, \apjs, 207, 3

\bibitem[{{Smith} {et~al.}(2017){Smith}, {Kilpatrick}, {Mauerhan}, {Andrews},
  {Margutti}, {Fong}, {Graham}, {Zheng}, {Kelly}, {Filippenko}, \&
  {Fox}}]{2017MNRAS.466.3021S}
{Smith}, N., {Kilpatrick}, C.~D., {Mauerhan}, J.~C., {et~al.} 2017, \mnras,
  466, 3021

\bibitem[{{Summa} {et~al.}(2013){Summa}, {Ulyanov}, {Kromer}, {Boyer},
  {R{\"o}pke}, {Sim}, {Seitenzahl}, {Fink}, {Mannheim}, {Pakmor},
  {Ciaraldi-Schoolmann}, {Diehl}, {Maeda}, \&
  {Hillebrandt}}]{2013A&A...554A..67S}
{Summa}, A., {Ulyanov}, A., {Kromer}, M., {et~al.} 2013, \aap, 554, A67

\bibitem[{{Veigele}(1973)}]{1973AD......5...51V}
{Veigele}, W.~J. 1973, Atomic Data, 5, 51

\bibitem[{{Weiler} {et~al.}(1998){Weiler}, {Van Dyk}, {Montes}, {Panagia}, \&
  {Sramek}}]{1998ApJ...500...51W}
{Weiler}, K.~W., {Van Dyk}, S.~D., {Montes}, M.~J., {Panagia}, N., \& {Sramek},
  R.~A. 1998, \apj, 500, 51

\bibitem[{{Wesson} {et~al.}(2015){Wesson}, {Barlow}, {Matsuura}, \&
  {Ercolano}}]{2015MNRAS.446.2089W}
{Wesson}, R., {Barlow}, M.~J., {Matsuura}, M., \& {Ercolano}, B. 2015, \mnras,
  446, 2089

\bibitem[{{Wheeler} {et~al.}(1975){Wheeler}, {Lecar}, \&
  {McKee}}]{1975ApJ...200..145W}
{Wheeler}, J.~C., {Lecar}, M., \& {McKee}, C.~F. 1975, \apj, 200, 145

\bibitem[{{Xu} {et~al.}(1992){Xu}, {McCray}, {Oliva}, \&
  {Randich}}]{1992ApJ...386..181X}
{Xu}, Y., {McCray}, R., {Oliva}, E., \& {Randich}, S. 1992, \apj, 386, 181

\end{thebibliography}

\label{lastpage}

\begin{appendices}

\section{The CSM Column Density} \label{ap.a}

The column density of absorbing particles is directly related to the equivalent width (EW) via the curve of growth, which has three regimes: linear, flat, and square root. D12 infer that the \ion{Ca}{2}~H\&K absorption lines are in the square-root regime based on their broad wings and large EW$_{\rm Ca~II}\approx 10$ \AA. However, if the lines are significantly broadened by the thermal and/or turbulent properties of the CSM, \ion{Ca}{2}~H\&K could be in the flat regime of the curve of growth, where a large EW is possible from a much lower column density of particles. In the flat regime, the expression for the relation between EW, column density, and line broadening is given by 

\begin{equation}\label{eq:Wflat}
{\rm EW} \approx \frac{2b\lambda}{c} \sqrt{ \ln{ \left( 1.13 \times 10^{20} \frac{c}{\pi^{0.5}} \frac{N\lambda f}{b} \right)} },
\end{equation}

\noindent
where EW is the equivalent width in \AA, $b$ is the line-broadening parameter in $\rm km\ s^{-1}$, $c$ is the speed of light in vacuum ($c = 3 \times 10^{5}$ $\rm km \ s^{-1}$), $f$ is the oscillator strength (unitless), and $N$ is the column density in $\rm cm^{-2}$. Assuming the true absorption-line profile is Gaussian, the line-broadening parameter $b$ is equivalent to the standard deviation and related to the line's FWHM by $b=$ FWHM/2.35. The maximum true width of the saturated \ion{Ca}{2} line in D12's high-resolution spectrum is $\rm{FWHM}\lesssim350$ $\rm km\ s^{-1}$, which implies $b\lesssim150$ $\rm km\ s^{-1}$. With Equation \ref{eq:Wflat} we find that the observed EW$_{\rm Ca~II} = 10$ \AA\ can result from a column density of just $N_{\rm Ca~II} = 10^{16}$ $\rm cm^{-2}$, more than two orders of magnitude lower than estimated by D12.

To check whether a line-broadening parameter of $b\lesssim150$ $\rm km\ s^{-1}$ is physically plausible for the PTF11kx system, we consider the two components of thermal and turbulent broadening. In Appendix A.1, we discuss information from the \ion{Na}{1} absorption features and find that a thermal component probably only accounts for $\lesssim 10\%$ ($b_{\rm therm} \approx 13$ $\rm km\ s^{-1}$) of the broadening. We propose that the remaining amount could be attributed to turbulence in the CSM, which is reasonable for shells of material created by a nova eruption sweeping up the companion's wind. For example, MB12 find that the coasting velocity of swept-up material $v_{\rm coast} \approx 300$ $\rm km\ s^{-1}$. The turbulent velocities would be $\sim30\%$ of this, leading to $b_{\rm turb}\approx 100$ $\rm km\ s^{-1}$. As a final estimate we also rederive the column density directly from optical depth in Appendix A.2, finding $N_{\rm Ca} \approx 10^{17}$ $\rm cm^{-2}$.

\subsection{Information from \ion{Na}{1}~D}\label{ap:a1}

In high-resolution Keck spectra obtained around the time of peak brightness, before the onset of interaction, presented by D12 (their Fig. 2), the \ion{Na}{1}~D 1 and 2 absorption features ($\lambda_{\rm NaID1}=5895.92$~\AA\ and $\lambda_{\rm NaID1}=5889.95$ \AA) show three components. All three appear to be unsaturated, and are at velocities $v_{\rm Na} \approx -65$, 0, and $+50$ $\rm km \ s^{-1}$ with respect to the host-galaxy reference frame. The EW of the blueshifted line increases after peak brightness, an effect that is commonly interpreted as the recombination of \ion{Na}{1} after an initial photoionization by the early-time SN's UV photons (e.g., \citealt{2007Sci...317..924P}). The blueshifted, saturated lines of \ion{Ca}{2}~H\&K are also centered at $v_{\rm Ca} \approx -65$ $\rm km \ s^{-1}$, and we know that they are from CSM because they transition to emission features upon interaction with the SN ejecta (D12). This implies a physical coincidence of the Na and Ca in the CSM. (As an aside, the two components of \ion{Na}{1}~D at  $v_{\rm Na} \approx 0$ and $+50$ $\rm km \ s^{-1}$ that do not change with time are likely generated by the host galaxy's interstellar material; from them we estimate the total column density of the interstellar material to be $N_{\rm Na} \approx 10^{12}$ $\rm cm^{-2}$ along the line of sight to PTF11kx).

The column density of recombined \ion{Na}{1} in the $v_{\rm Na} \approx -65$ $\rm km \ s^{-1}$ component at $\sim20$ days after peak is $N_{\rm Na} \approx 10^{12}$ $\rm cm^{-2}$. In \S~\ref{ssec:disc.N} we show that the column density of calcium is $N_{\rm Ca} \gtrsim 10^{16}$ $\rm cm^{-1}$, and since calcium and sodium have similar number fractions in solar-abundance material, this suggests that only a very small fraction of the sodium has recombined. This is to be expected because the ionization energies for \ion{Na}{1} and \ion{Ca}{1} are quite similar, $E_{\rm NaI}=5.14$ $E_{\rm CaI}=6.11$ $\rm eV$, but the recombination timescale for \ion{Ca}{2} is longer. Similar observations of \ion{Na}{1}~D recombination and a nonevolving \ion{Ca}{2} feature were presented for SN\,Ia 2006X by \cite{2007Sci...317..924P}. In fact, if the CSM environment is $T \approx 10^4$ $\rm K$ as predicted by MB12, most of the sodium and calcium would be ionized even without the influence of the SN\,Ia radiation. With only a small amount of the sodium in the ground state and available for photoionization by the young SN's UV photons, only a small amount of recombination is expected. 

This small amount of recombined \ion{Na}{1} represents the CSM that is hot but perhaps not very turbulent (otherwise it may be collisionally ionized). This feature is unsaturated and has a line width of FWHM $\approx 40$ $\rm km \ s^{-1}$, or $\Delta \lambda_{\rm FWHM} \approx 0.8$ \AA, which is related to the temperature of the emitting material under the assumption of thermal broadening by 

\begin{equation}
\Delta \lambda_{\rm FWHM} = \sqrt{ \left( \frac{8k_B T \ln{2}}{m c^2} \right) }  \lambda_0,
\end{equation}

\noindent
where $k_B$ is the Boltzmann constant ($k_{\rm B} = 1.38 \times 10^{-16}$ $\rm erg \ K^{-1}$), $T$ is the temperature of the line-forming material,  and $m$ is the atomic mass of the element creating the line ($m_{\rm Ca} = 40$ and $m_{\rm Na}=23$ $\times 1.66 \times 10^{-24}$ $\rm g$). Based on the observed line width of \ion{Na}{1}, we estimate that $T\approx 8 \times 10^5$ $\rm K$ in the recombined \ion{Na}{1}, which we note is hotter than predicted for wind material swept up into shells of CSM by MB12, $T \approx 10^4$ $\rm K$. We then use this same equation to estimate the thermal broadening of \ion{Ca}{2} at this temperature, and find FWHM $\approx 30$ $\rm km \ s^{-1}$. Assuming the absorption feature is Gaussian, the component of the standard deviation from thermal broadening in the \ion{Ca}{2} line is $b_{\rm therm} \approx 13$ $\rm km \ s^{-1}$.

\subsection{Direct Derivation from Optical Depth}\label{ap:a2}

The specific intensity ($I_\nu$) in an absorption line as a function of frequency ($\nu$) is given by 
\begin{equation}
I_{\nu} = I_0 (\nu) e^{-\tau(\nu)},
\end{equation}
\noindent where $I_0$ is the continuum (or pseudocontinuum, the flux level on either side of the absorption line) and $\tau(\nu)$ is the optical depth. The optical depth is related to the column density, $N$, of the absorbing material and its absorption cross section, $\sigma(\nu)$, via
\begin{equation}
\tau(\nu) = N \sigma(\nu);
\end{equation}
\noindent
thus, we can derive the column density directly from the 
\ion{Ca}{2} line without using curve-of-growth relations if we know $\sigma(\nu)$. The absorption cross section of an atomic line is given by
\begin{equation}\label{eq:sigmanu}
\sigma(\nu) = \frac{\pi q_e^2}{m_e c} f_\mathrm{osc} \phi(\nu) 
            = (1.7\times10^{-2}~\mathrm{cm^2~Hz})~ \phi(\nu),
\end{equation}
\noindent where $q_e$ is the electron charge, $m_e$ is the electron mass, 
$f_\mathrm{osc}$ is the line oscillator strength (0.682 for the \ion{Ca}{2} K line), and $\phi(\nu)$ is the line profile. We assume the profile is Gaussian owing to thermal (and turbulent) broadening of a Lorenzian:
\begin{equation}\label{eq:phinu}
\phi(\nu) = \frac{1}{\sqrt{\pi} \Delta\nu} 
            \exp\left[ -\left( \frac{\nu - \nu_0}{\Delta\nu}\right)^2\right],
\end{equation}
\noindent
where $\Delta\nu$ is the line width and $\nu_0$ is the central line frequency 
($\nu_\mathrm{Ca~II,K}=7.63\times10^{14}~\rm Hz$). 

In the spectrum we look for the frequency $\nu_1$ at which $\tau = 1$ (i.e., $I_{\nu} = 0.37\,I_0$) 
and thus $\sigma(\nu_1) = N^{-1}$. 
For the saturated \ion{Ca}{2} line in D12, 
$\tau=1$ at $\sim 400$ $\rm km\ s^{-1}$, so 
$\nu_1 \approx \nu_{\rm Ca} + 10^{12}~\rm Hz$. 
Using FWHM $\lesssim 350~ \rm km~ s^{-1}$, as in \S~\ref{ssec:disc.N}, 
the line width is $b = {\rm FWHM}/2.35$, which when translated into frequency 
gives $\Delta \nu \lesssim 3.8\times10^{11}~\rm Hz$. 
With these numbers, we find that $\phi(\nu_1) = 1.4\times10^{-15}~\rm Hz$, 
$\sigma(\nu_1) = 2.3 \times 10^{-17}~\rm cm^2$, 
and that the column density of \ion{Ca}{2} is $N_{\rm Ca~II} \approx 4\times10^{16}$ $\rm cm^{-2}$. 
This is lower than the column density estimate of D12 (assuming the square-root regime of the curve of growth), and slightly higher than our own estimated lower limit assuming the 
flat regime of the curve of growth presented above. 

\end{appendices}

\end{document}